\documentclass[a4paper,10pt]{article}
\pdfoutput=1
\usepackage{graphicx}
\usepackage{epsfig}
\usepackage{bm}
\usepackage{amsmath}
\usepackage{amssymb}
\usepackage{graphics}
\usepackage{color}
\RequirePackage{slashed}
\usepackage[linkcolor=red,citecolor=green,urlcolor=blue]{hyperref}
\usepackage{appendix}

\usepackage{soul}
\newcommand\ddfrac[2]{\frac{\displaystyle #1}{\displaystyle #2}}

\usepackage{booktabs}
\usepackage{comment}
\usepackage{authblk}
\newcommand{\ie}{\textsl{i.e. }}
\newcommand{\eg}{\textsl{e.g. }}
\newcommand{\bp}{b_\perp}
\newcommand{\si}{\sigma_{eff}}
\usepackage[normalem]{ulem} 
  
  \usepackage[a4paper,top=2cm,bottom=2cm,left=2cm,right=2cm]{geometry}

\date{}
\title{Double parton scattering and the proton transverse structure at the 
LHC}

\author[1]{Matteo Rinaldi}
\affil[1]{Dipartimento di Fisica 
e Geologia, Universit\`a degli studi di Perugia and \newline 
INFN Sezione di Perugia, Via A. Pascoli, I-06123 Italia}

\author[2,3]{Federico Alberto Ceccopieri}

\affil[2]{D\'ep. AGO, Universit\'e de Li\`ege and IFPA, all\'ee du 6 
Aout, 17
4000 Li\`ege 1, Belgium }
\affil[3]{ Department of Physics, Technion – Israel Institute of 
Technology, 
Haifa, 32000 Israel}

\begin{document}
 
\maketitle
 
 \begin{abstract}
 \noindent
 We consider double parton distribution functions (dPDFs), essential 
 quantities in double parton scattering (DPS) studies,
which encode novel non perturbative insight on the partonic proton 
structure. We develop the formalism to extract this information from 
dPDFs and present results by using constituent quark model  calculations 
within the Light-Front approach,   
focusing on radiatively generated gluon dPDFs. 
Moreover, we generalize the relation between the mean transverse 
partonic distance between two active partons in a DPS process and   the 
so called $\sigma_{eff}$ to include partonic correlations and the so 
called 2v1 mechanism contribution.
Finally we investigate the impact of relativistic effects on digluon 
distributions and study 
the structure of the corresponding
longitudinal and transverse
 correlations.

\end{abstract}


\maketitle
\section{Introduction}
\label{intro}
A proper description of the event structure in hadronic collisions requires the 
inclusion of the so called multiple parton interactions (MPI) which affect both 
the   multiplicity and  structure of the hadronic final 
state \cite{paver,Sjostrand:1986ep}. 
The Large Hadron Collider operation 
renewed the interest in MPI given the continuous demand for an increasingly 
detailed description of the hadronic final state which is crucial in many 
New Physics searches.
In this rapidly evolving context, these type of studies have also received
attention 
for their own sake since they might be sensitive to double partonic
correlations in the 
colliding hadrons, see recent review in Ref. \cite{Kasemets:2017vyh}. 
The simplest MPI process is the double parton scattering (DPS) 
\cite{Goebel:1979mi,Mekhfi:1983az}.
In such a process, a large momentum transfer is involved  in both
scatterings which enables the use of perturbative techniques 
to calculate the corresponding cross 
section. The latter depends  on two-body 
quantities,  the so called double Parton Distribution Functions 
(dPDFs), which are interpreted as the  number densities of  parton pair 
at a given  transverse distance, $b_\perp$,  and carrying  longitudinal 
momentum fractions 
($ x_1,x_2$) of the parent proton \cite{paver,ww_3_1,Calucci:1999yz}.
Double PDFs are not  perturbatively calculable from first principles, 
a feature shared with usual PDFs and other quantities in QCD. Moreover, due to 
their  
dependence upon the partonic interdistance \cite{Calucci:1999yz},
they contain information on the hadronic structure    
complementary to those obtained from one-body distributions such as
generalized parton distribution functions (GPDs) and transverse momentum 
dependent PDFs (TMDs).
Unfortunately the DPS cross section is obtained by integrating
dPDFs over $b_\perp$ so that such a dependence is not directly measurable 
\cite{paver}. 
In this scenario, hadronic models have been used to 
obtain basic information on dPDFs and to gauge the phenomenological impact of
longitudinal and transverse 
correlations, see Refs. 
\cite{Mel_19,noiold,Mel_23,noi1,noir,noipion}, along with spin correlations 
\cite{noi1,cotogno,muld,positivity,Kasemets:2012pr}.
We mention that  quantities 
{related} to dPDFs, and encoding double parton correlations, have been 
recently calculated for pion by means of Lattice techniques \cite{lattice}.
Despite the wealth of information encoded in dPDFs, present {experimental} 
knowledge on 
DPS cross section is accumulated, up to now, into the so called 
effective cross section, $\sigma_{eff}$, for recent results see \eg  
Refs. \cite{data6,data8,data9,data10,data11,data12}.  
The latter is defined through the ratio of the product of  two single 
parton scattering 
cross sections to the DPS cross section with the  same final states. 
In the present paper, we
continue the investigation of the relationship between $\sigma_{eff}$ and 
the mean interpartonic 
distance pursued in Ref. \cite{rapid}. We study its modification 
induced by including
the so called 
splitting 2v1 term contribution in DPS processes \cite{ww_23,Ceccopieri:2010kg,ww_25_1,Gaunt:2009re,gauntladder,Treleani:2018dbg,
Ryskin:2011kk,73,74,77}  
and, separately, the effects 
of longitudinal correlations in dPDFs.
Numerical estimates will be shown and discussed in the kinematics 
of DPS processes 
initiated by digluon distributions, see e.g. Refs. 
\cite{mulders,Golec-Biernat:2015aza,Elias:2017flu,dj,Golec-Biernat:2014bva} 
on this topic, 
which are perhaps the most interesting distributions in the DPS context. Such 
distributions are radiatively generated by pQCD 
evolution \cite{ww_3_1,ww_23,Ceccopieri:2010kg,ww_25_1,Ryskin:2011kk,73,74,77,
Diehl:2015bca,LH3,75,keane,snigireva} starting from 
{valence} dPDF model calculation at the  hadronic scale, $Q_0$. The digluon
distrubution, in principle, is likely to have 
a non perturbative contribution at $Q_0$. 
In the present work,  we make use of a pure radiative evolution scheme, and 
therefore
it is our precise choice to neglect such an additional input which 
requires the  
modelization of the
non perturbative sea quarks and gluons distributions, such those proposed in
Ref. \cite{noij2}.
{On the other hand it must be emphasized  
that in case of ordinary DIS structure functions measurements, 
predictions based on parton 
distributions evolved in such a scheme
are not able to  reproduce the small $x$ behaviour of the data and a non perturbative sea quarks and gluon PDFs input are required.} 
Moreover, in such a radiative scheme and given the structure of dPDFs evolution 
equations, digluon interdistance 
follows the pattern 
of that of valence quarks 
obtained
 from  the underlying hadronic 
model used for the dPDF calculations. 
DPS measurements, sensitive to gluon initiated processes, will then provide a
test of our  approach.  
In the last part of the paper, we focus on relativistic effects in dPDF 
model calculations, in the relevant kinematic conditions of collider 
experiments, and already addressed in Ref.  \cite{noir} for valence 
quarks  at the hadronic scale. 
{The aim of this part of the analysis is to offer insight to unfactorized ansatz for dPDFs as induced by the implementation  of relativistic effects in dPDFs calculation via Melosh operators. }

The paper is organized as follows.  In Sec.  \ref{process} we will discuss 
processes and corresponding kinematic conditions which we will focus upon in 
this analysis.
In Sec. \ref{proton} we describe the  formalism which allows to obtain
physical information on the proton structure  from dPDFs. In Sec. \ref{si1} we 
introduce the 
so called $\sigma_{eff}$, relevant quantity  in DPS analyses and show how 
the latter is related to the geometrical properties of the proton.
In Sec. \ref{mel} we discuss relativistic effects in dPDF calculations 
and then present our Conclusions.

\section{Analysis strategy and calculation details}
\label{process}
\noindent
In the present analysis 
we will focus on the digluon distributions
and therefore we consider DPS prototype processes: 
\begin{equation}
pp \rightarrow J/\Psi J/\Psi X, \;\;\; 
pp \rightarrow H H X\;, 
\label{processes}
\end{equation}
{for which the production mechanism is dominated by gluons} and 
where each final state particle is produced 
in a distinct parton-parton scattering.
Double $J/\Psi$ production has been already measured both at Tevatron and 
LHC \cite{data8,data11,data12,data14,lans}.  Double Higgs production via 
DPS has been 
studied in the literature \cite{gauntladder}, but not yet measured  given its 
rather low cross section.
We mention here that it would be also interesting to consider the mixed process
$pp \rightarrow H J/\Psi X$ with final state 
produced via DPS which,  to the
best of our knowledge, has never been
considered in the literature. 
We also mention that interesting information could be gained from the 
comparison of  the double $J/\Psi$ production with the double open charm one in the same kinematics.
The combined measurements of these DPS processes,
among many others with a less pure gluonic initial state but larger 
cross sections, give a wide coverage of digluon
distribution both in hard scale and fractional momenta. 
We define  the partonic subprocess
in the two scatterings in Eq. (\ref{processes}) as 
\begin{equation}
p_i(k_1)+p_k(k_3) \rightarrow A(k_A) X \;\; \mbox{and} \; \; 
p_j(k_2)+p_l(k_4) \rightarrow B(k_B) X, \;\; \mbox{with} \;\; A,B=J/\Psi,H\,,
\label{processes2}
\end{equation}
where $p$'s and $k$'s are the relevant parton flavour 
and momenta, respectively.
Since heavy particles appearing in Eq. 
(\ref{processes}) are produced by partonic annihilation
in lowest order of perturbation theory, the fractional 
momenta  of the incident gluons can be reconstructed 
from the mass $m$, transverse momentum
 $k_T$ and rapidity $y$ of final state particles as 
\begin{equation}
x_{1,3}=\frac{\sqrt{m_A^2+k_{T,A}^2}}{\sqrt{s}} e^{\pm y_A}, \;\;\;
x_{2,4}=\frac{\sqrt{m_B^2+k_{T,B}^2}}{\sqrt{s}} e^{\pm y_B}\,. 
\end{equation}
In our calculations we set the centre-of-mass energy to its 
nominal value at the LHC, 
$\sqrt{s}$=13 TeV, and consider two rapidity region: the central
one covered by ATLAS and CMS, $|y| < 1.2$ and the forward one covered by LHCb, 
$2<y<4.5$.
Neglecting transverse momentum, 
$J/\Psi$ production gives access 
to  fractional momenta in the range 
$ 10^{-6} \lesssim x \lesssim 10^{-2}$ while Higgs production  in 
the range $10^{-4} \lesssim x \lesssim 1$. The factorization scale in each 
process is set equal to the mass of the particle, either the $J/\Psi$ or Higgs 
boson, produced in the final state, 
$\mu_{F,A}=m_A$ and $\mu_{F,B}=m_B$ with
$m_{J/\Psi}=2m_c$.
The differential DPS cross section, assuming that the two hard scatterings 
can be factorized 
\cite{ww_3_1,ww_25_1,Diehl:2015bca,Gaunt:2014ska,add1,Diehl:2018wfy}, involves 
dPDFs through an integral over the transverse partonic distance $b_\perp$ and 
reads \cite{paver,ww_3_1}: 
\begin{equation}
 d\sigma_{DPS}^{A+B}= \frac{m}{2}  \int
 d^2b_{\perp}
d\hat{\sigma}_{ik}^A 
\; d\hat{\sigma}_{jl}^B \; \tilde F_{ij}(x_1,x_2,b_{\perp})
 \tilde F_{kl}(x_3,x_4,b_{\perp})
 \,.
 \label{defi}
\end{equation}
In Eq. (\ref{defi}) 
 $d\hat{\sigma}$ are the differential partonic cross
 sections for processes with final state A or B
 respectively and the symmetry factor reads $m=1$ if $A=B$
 and $m=2$ otherwise.
Double PDFs appearing in Eq. (\ref{defi}), are multidimensional distributions 
encoding non perturbative features of the proton structure and are therefore 
complicated to model. Some guidance in building 
appropriate initial conditions is offered by physical intuition at small $x$ 
\cite{ww_3_1,ww_25_1,Gaunt:2009re,Diehl:2015bca,keane,add2}  
and by sum rules 
\cite{Gaunt:2009re,Golec-Biernat:2015aza,Golec-Biernat:2014bva,Ceccopieri:2014ufa}.
Nevertheless, a large freedom is left in the gluon transverse 
spectrum, which is perhaps one of the most intriguing aspect for hadronic 
studies. In order to investigate some of these features,
in the present paper we make use of  dPDF
calculations within constituent quark models (CQMs),  e.g. Refs. 
\cite{Mel_19,noiold,Mel_23,noi1}. Following  
the line of Ref. \cite{noir}, we have adopted the  hypercentral quark 
model (HP), in its relativistic version \cite{faccioli}
and, in order to highlight model independent effects on dPDFs, 
the harmonic oscillator model (HO) \cite{giannini}.
In particular, for the latter, we 
considered the version described in Ref. \cite{noir}, where the model parameter 
$\alpha$,  representing the width of the Gaussian, is set to be  $\alpha^2 = 
25$ fm$^{-2}$ in order to mimic a relativistic structure.
These models differ from each other in many dynamical aspects and offer
a parametrization of the only non-vanishing valence-valance dPDF
at the hadronic scale, $Q_0$. All other distributions are then radiatively
obtained at higher scales  by performing pQCD evolution in its homogeneous 
form, which is appropriate at fixed $b_\perp$ 
\cite{ww_3_1,keane,Diehl:2018kgr}. 
{The value of the hadronic scale $Q_0$  
has been fixed according to the procedure outlined  in Ref. \cite{noiww}, 
\textsl{i.e.} by tuning its value in order to reproduce known SPS cross sections 
by using single PDFs obtained by the same hadronic model
and evolved starting from $Q_0$.
The obtained value is given by $Q_0^2=0.26$ Ge$\mbox{V}^2$.
Since both single and double PDFs are built upon  the same hadronic model, 
$Q_0$ is used also as starting scale for dPDFs evolution}. Since 
$Q_0$ is located  in the infrared region, 
both distributions  
show a large sensitivity to its precise value.
In order to reduce the impact of this choice on our results, we will often 
consider appropriate ratios
involving single and double PDFs which decrease, and in many cases almost 
cancel, this dependence. 
This feature is 
particularly relevant for the calculation of the effective cross section 
which we will be introduced in the next Section.


\section[Proton structure from dPDFs]{Proton transverse structure from dPDFs}
\label{proton}
In this section we  present the general formalism necessary to extract physical information on the proton structure  from dPDFs, 
\textsl{i.e.} the mean partonic distance between two partons in the transverse 
plane. 
These results are completely general and do not require any specific assumption on dPDFs.
Since the latter represent the number density of two parton with 
longitudinal momentum fractions $x_1$ and $x_2$ at a given transverse 
distance $b_\perp$ \cite{paver}, they 
provide a new tool to access the 3D structure of the proton, complementary to that obtained from generalized parton distribution functions (GPDs).
In particular, these two-body functions are sensitive to double parton 
correlations 
\cite{Kasemets:2017vyh,Mel_19,noiold,Mel_23,noi1,noir,noipion,cotogno,
lattice,77,mulders,keane,snigireva,noij2} 
that can not be accessed by means of one-body distributions such as GPDs. 
To this aim we first introduce the effective 
form factor (EFF) \cite{noiplb1,rapid} as
discussed in Ref. \cite{rapid}, \textsl{i.e.} by means of the hadron wave
function $\Psi$ in the non relativistic limit:
\begin{align}
\label{uno}
 f_{ij}(k_\perp)= \int d\vec k_1 d\vec k_2~ \Psi^\dagger(\vec k_1+ 
 \vec k_\perp, \vec k_2) \tau_i \tau_j
 \Psi(\vec k_1, \vec k_2 + \vec k_\perp)~,
\end{align}
where $\vec k_i$ is the total momentum of the parton $i$ and $\tau_i$ 
the standard flavor projector. 
As discussed in Ref. \cite{rapid},  $k_\perp$ 
represents a transverse momentum imbalance between 
two partons in the amplitude and its conjugate \cite{ww_25_1}.
The EFF represents the Fourier Transform of the number distribution of
two partons at a given transverse distance \cite{rapid,noiplb1}:
\begin{align}
  f_{ij}(k_\perp) = \int d^2 b_\perp ~e^{\vec k_\perp \cdot \vec b_\perp} 
\tilde f_{ij}(b_\perp).
 \label{fteffff}
\end{align}
This distribution can be written in terms of dPDFs in coordinate space, 
\textsl{i.e.} $\tilde F_{ij}(x_1,x_2,b_\perp)$:
\begin{align}
 \tilde f_{ij}(b_\perp) = \int dx_1~dx_2~\tilde F_{ij}(x_1,x_2,b_\perp)~.
\end{align}
$\tilde F_{ij}(x_1,x_2,b_\perp)$ encode 
information on the
proton structure such
as correlations between 
the longitudinal momentum 
fractions of two partons and their partonic distance. 
The latter, for a pair of partons with flavour $i$ and
$j$ and fractional momenta $x_1$ and $x_2$, is defined as
\begin{align}
 \langle b_\perp^2\rangle^{ij}_{x_1,x_2} & =\ddfrac{\int d^2b_\perp  ~ b^2_\perp 
\tilde F_{ij}(x_1,x_2,b_\perp, Q^2)}{\int d^2b_\perp ~ \tilde 
F_{ij}(x_1,x_2,b_\perp,Q^2) }~,
\label{dx4}
\end{align}
where $Q^2$ is a  generic hard  scale at which dPDFs
are evaluated and we have denoted $b_\perp \equiv |\vec b_\perp|$.
It is easy then to show that 
the mean partonic distance averaged
over parton fractional momenta is 
given by
\begin{align}
    \langle b_\perp^2 \rangle^{ij} = \ddfrac{ \int d^2b_\perp ~b_\perp^2 \tilde 
f_{ij}(b_\perp)}{\int d^2b_\perp ~ \tilde f_{ij}(b_\perp)}~.
\end{align}
{The above quantities can be related to each other as follows:}

\begin{align}
\label{probab}
    \langle b_\perp^2 \rangle^{ij} &= \int dx_1~dx_2 ~\langle b_\perp^2 
\rangle_{x_1,x_2}^{ij} P_{ij}(x_1,x_2)~,
\end{align}
where $P_{ij}(x_1,x_2)$ represents the probability of finding a pair of partons 
with flavours $i,j$ and longitudinal momentum fractions $x_1,x_2$:
\begin{align}
    P_{ij}(x_1,x_2) = \ddfrac{\int d^2 b_\perp~\tilde 
F_{ij}(x_1,x_2,b_\perp)}{\int dx_1~dx_2~\int d^2 b_\perp~\tilde 
F_{ij}(x_1,x_2,b_\perp)}~.
\end{align}

%
%
%
As for the standard electro-magnetic nucleon form factor, 
such a   relation can be equivalently obtained from
dPDFs in momentum space, \textsl{i.e.}
 $F_{ij}(x_1,x_2,k_\perp)$,  the Fourier transform (FT) of the dPDF 
$\tilde F_{ij}(x_1,x_2,b_\perp)$ in coordinate space.
{Likewise}, as for GPDs,  $F_{ij}(x_1,x_2,k_\perp)$ does not 
admit a 
probabilistic interpretation in $k_\perp$-space, which holds
instead in $b_\perp$-space. Since
\begin{equation}
\label{dd1}
F_{ij}(x_1,x_2,k_\perp;Q^2) =
 \int d^2b_\perp e^{i \vec b_\perp \cdot \vec 
k_\perp}\tilde F_{ij}(x_1,x_2,b_\perp,Q^2) \sim
 \int d^2b_\perp \left( 1- {1 \over 4} k_\perp^2 
b_\perp^2 \right) \tilde F_{ij}(x_1,x_2,b_\perp;Q^2)~, 
\end{equation}
it follows that 
\begin{equation}
 \int d^2b_\perp b_\perp^2 \tilde F_{ij}(x_1,x_2,b_\perp;Q^2) =
-4{d \over 
dk_\perp^2}F_{ij}(x_1,x_2,k_\perp;Q^2) \Big|_{k_\perp=0}~.
\label{new}
\end{equation}
From the above relation, Eq. (\ref{dx4})  can be equivalently 
written in terms of dPDFs in momentum space, in
analogy with the standard electro-magnetic form
factor:
\begin{align}
 \langle b^2 
\rangle_{x_1,x_2}^{ij} = -4 {d \over d k_\perp^2}  
\left[ {F_{ij}(x_1,x_2,k_\perp;Q^2) 
\over 
 F_{ij}(x_1,x_2, k_\perp=0;Q^2)   }   \right]_{k_\perp=0}~\,.
 \label{dx3}
\end{align}
{Given the really limited knowledge on dPDFs driven by data, one can explore this approach 
by using dPDFs obtained from hadronic model calculations.}
 \begin{figure*}[t]
\vskip 0.5cm
\hskip 0.5cm
\includegraphics[scale=0.7]{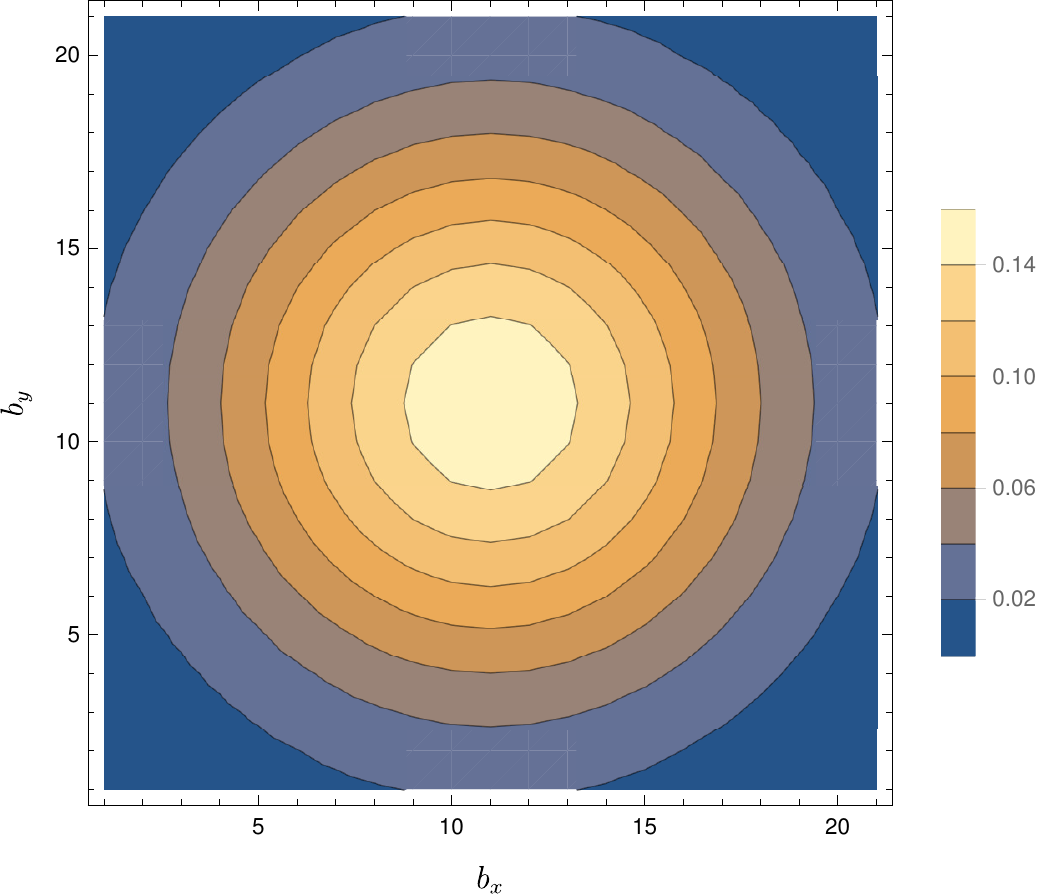}
\hskip 0.5cm
\includegraphics[scale=0.7]{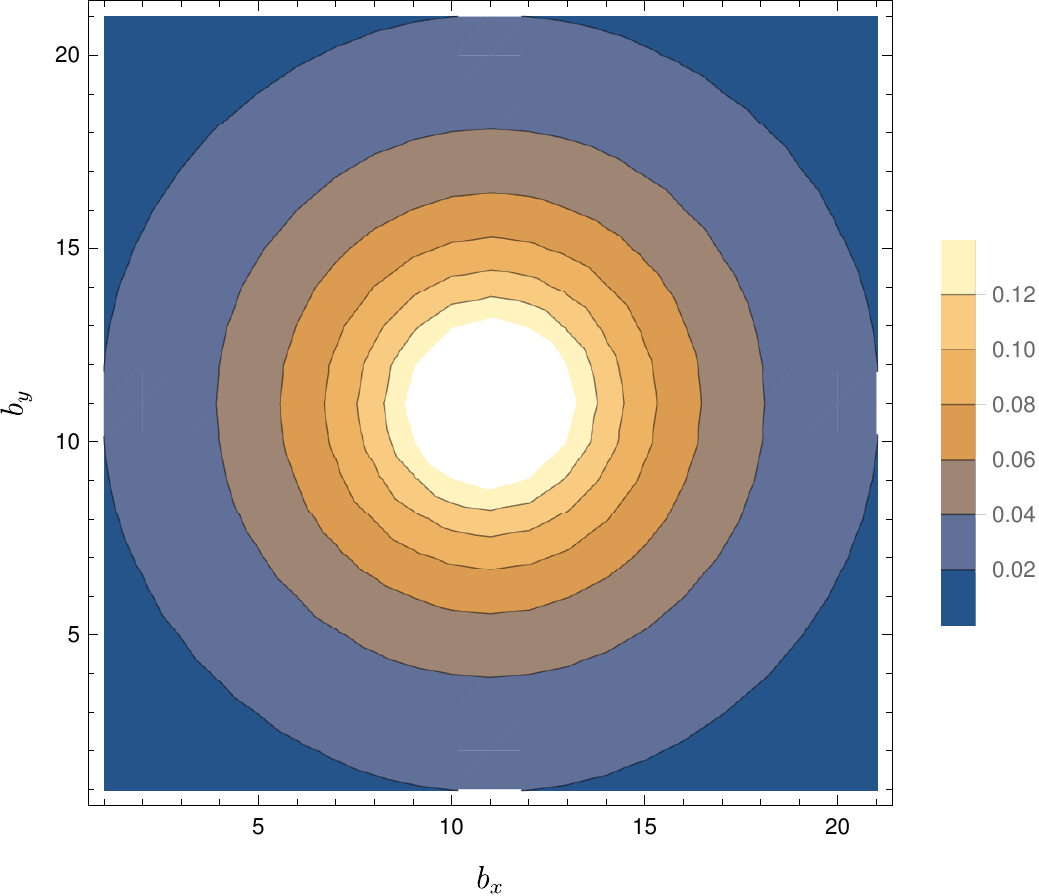}
\caption{ \textsl{The digluon distribution $\tilde 
F_{gg}(x_1=10^{-4},x_2=10^{-2},b_\perp,Q^2=m_H^2)$. 
 Left panel:  
calculation within the HO model. Right panel: 
calculation within the HP model. Partonic distance expressed in [GeV$^{-1}$]. }}
\label{tomo3}
\end{figure*}
In Fig. \ref{tomo3} we present the digluon dPDFs, 
evaluated within the HO (left panel) and HP (right 
panel) models at $Q^2=m_H^2$ in $\vec b_\perp$-space. Since we consider 
unpolarized partons in an unpolarized proton,  
circular symmetry in $\vec b_\perp$ is obtained, as apparent from the 
plot. 
Furthermore, the shape  of the distributions are 
qualitatively similar to those shown 
in Ref. \cite{noir}, where valence quark dPDFs have 
been evaluated within the same models but at the 
hadronic scale. 
By using these quantities,
we have also evaluated the mean gluonic 
distance via Eqs. (\ref{dx4},\ref{dx3}). The results, 
reported
in  Tab. \ref{tablebb}, show that 
partonic correlations induce a dependence of 
the mean partonic distance 
upon the longitudinal momentum fractions carried by 
two partons. We recall that 
if correlations between $x_i$ and $k_\perp$ were 
absent, as in the non 
relativistic limit of dPDFs evaluated within the HO 
model (see Ref. 
\cite{noiold}), the mean partonic distance  would not 
depend on $x$'s and reads $\sqrt{\langle b^2 
\rangle}=0.283$ fm.
\begin{table}[t]
\begin{center}
\centering
\begin{tabular}{|c|c|c|}
\hline
Kinematics &  HO model & HP model \\ 
$x_1,x_2$  &  $\sqrt{\langle b^2 \rangle_{x_1,x_2}}$ 
[fm] & $\sqrt{\langle b^2 \rangle_{x_1,x_2}}$ [fm] \\ 
\hline
$10^{-4},10^{-4}$  & 0.393 & 0.391\\
$10^{-4},10^{-2}$ & 0.382 & 0.408\\
$10^{-4},0.4~~$ & 0.393 & 0.405\\
$10^{-3},10^{-3}$  & 0.383 & 0.407\\
$10^{-2},10^{-2}$  & 0.365 & 0.404\\
$10^{-2},0.4~~~$  & 0.377 & 0.377\\\hline
\end{tabular}
\caption{ \textsl{Mean intergluon distance evaluated via dPDFs 
calculated at the scale 
$Q^2=m_H^2$
with the HO and HP models in different $x_1,x_2$ configuration.}}
\label{tablebb}
\end{center}
\end{table}
This discussion, however, is rather academic since 
the present accuracy of DPS measurements is far 
from being sensitive to this kind of effects. 
Nevertheless we have shown in Ref. 
\cite{rapid} that physical information on the 
proton structure can still be directly obtained  from 
$\sigma_{eff}$, a quantity which is often used in 
experimental analyses. 
In next sections we review the formalism
that allows one to relate $\sigma_{eff}$ to $\langle 
b^2 \rangle$, and generalize it  to more complicated 
cases.

\section{ Transverse Proton structure from effective 
cross section}
\label{si1}
{Double PDFs, the main non-perturbative 
ingredients appearing in the cross section formula in Eq. (\ref{defi}),  are basically unknown, 
so that the direct application of the methods outlined in the previous Section can not be presently used. 
In this Section we  discuss
an alternative method that allows us to obtain information on the proton structure starting from experimental extracted quantities such as $\sigma_{eff}$.
We proceed in the analysis with an increasing degree of complexity:
in the first part of the Section,
we find useful to summarize the strategy of the evaluation for the most simple case, \textsl{i.e.} a fully factorized ansatz of dPDFs \cite{rapid}. 
In the second part, we generalize the results to include the so called splitting contribution to dPDFs which embodies correlations of perturbative origin. In the third part we generalize these results to unfactorized ansatz for dPDFs. In the last part of the Section all these results have been combined in a fully general relation between $\sigma_{eff}$ and $\langle b^2 \rangle_{x_1,x_2}$.}

\subsection{The factorized case}
\label{ss1}
In Ref. \cite{rapid} we have derived a relation between $\sigma_{eff}$ and the mean transverse partonic distance within the most simple assumptions on dPDFs, 
the fully factorized ansatz: 
\begin{align}
F_{ij}(x_1,x_2,k_\perp) \sim 
q_i(x_1)q_j(x_2)f(k_\perp)\,,
\label{ans}
\end{align}
where $q_i(x)$ are ordinary single PDFs  and 
$f(k_\perp)$ is the effective form factor defined in 
Eq. (\ref{uno}). Usually,  in such a simplified 
approach, 
$f(k_\perp)$ does not depend 
 on the parton flavors {nor on fractional momenta}
 \cite{Mekhfi:1983az,gauntladder}.
These assumptions allows  to rewrite the DPS cross 
section as \cite{Calucci:1999yz,trele}
\begin{equation}
 \label{sff}
 d\sigma_{DPS}^{A+B}  = 
 \frac{m}{2}\dfrac{d\sigma_{SPS}^A \, d\sigma_{SPS}^B
}{ \sigma_{eff}}~,
\end{equation}
being $d\sigma_{SPS}^{A(B)}$ the single parton 
scattering cross sections with
final state $A(B)$. In this 
scenario $\sigma_{eff}$  simply reads:
\begin{align}
\label{sap}
 \sigma_{eff}^{-1} = \int { d^2k_{\perp} \over 
 (2\pi)^2} f(k_\perp)^2~
{=\int { d k_{\perp} \over 2\pi} k_{\perp} 
f(k_\perp)^2} ,
\end{align}
where the last expression follows from rotational invariance. Eq. 
(\ref{sff}) shows that, in such an approximations, $\sigma_{eff}$
enters the DPS cross section as an overall 
normalization factor. 
We remark that the EFF entering in the above is defined similarly to 
that in Eq. (\ref{uno}) but without the partonic flavor dependence, as often assumed in the experimental analyses in which $\si$ is extracted.
In Ref. \cite{rapid}, we have shown that, by 
using the formal definition of the EFF in Eq. (\ref{uno}) and appearing in 
Eq. (\ref{sap}), one can relate $\sigma_{eff}$ to the mean partonic 
distance of two partons active in a DPS process. We will briefly review 
this procedure in the following.
As discussed in e.g. Refs. \cite{noiplb1,gauntladder,rapid,noir}, the EFF is the FT of the probability distribution of finding two 
parton at a given transverse distance, \textsl{i.e.} $\tilde f(b_\perp)$, in a confined  quantum mechanical
system:
\begin{align}
  f(k_\perp) = \int d^2 b_\perp ~e^{\vec k_\perp \cdot \vec b_\perp} 
\tilde f(b_\perp).
 \label{fteff}
\end{align}
In terms of the latter, $\si$ Eq. (\ref{sap}) is simply given by: 
\begin{align}
 \sigma_{eff}^{-1} = \int d^2\bp~ \tilde f(\bp)^2~,
\end{align}
see e.g. Refs. \cite{gauntladder,trele,Gaunt:2009re}. 
The latter expression relies on the probabilistic interpretation of $\tilde 
f(\bp)$:  this quantity represents the probability of finding a pair of 
partons at transverse distance $\bp$ \cite{gauntladder,trele,Gaunt:2009re}. This condition imposes the following normalization:
\begin{align}
 \int d^2 b_\perp ~ 
\tilde f(b_\perp)=1~.
\end{align}
This is a common assumption used in many phenomenological analyses of $\si$, see 
e.g. Ref. \cite{gauntladder}. The probabilistic interpretation of $\tilde f(\bp)$
is transparent, for example, in the 
non relativistic limit. 
In fact, by considering Eq.(\ref{uno}), one gets:
\begin{align}
 \tilde f(\bp) &= \int \ddfrac{d^2k_\perp}{(2 \pi)^2} ~e^{-i \vec k_\perp \cdot 
\vec b_\perp}f(k_\perp)
\\
\nonumber
&= \int \ddfrac{d^2k_\perp}{(2 \pi)^2} ~e^{-i \vec k_\perp \cdot 
\vec b_\perp} \int d\vec k_1 d\vec k_2~\Psi^\dagger(\vec k1+\vec k_\perp, \vec 
k_2) \Psi(\vec k_1,\vec k_2+\vec k_\perp)
\\
\nonumber
&\propto \int d\vec b_1 d\vec b_2 ~ | \tilde \Psi(\vec b_1,\vec b_2) |^2 
\delta^{(2)}(\vec b_{2\perp}-\vec b_{1 \perp}- \vec b_\perp)~,
\end{align}
where here $\tilde \Psi(\vec b_1,\vec b_2)$ is the proton wave function in 
coordinate space and $b_i$ is the position of the parton $i$ in center mass frame. In terms of the EFF, 
two asymptotic conditions, related to the above features,  can be obtained 
similarly to the standard 
form factors:  
\begin{align}
f(k_\perp=0)=1 \; \; \mbox{and} ~f(k_\perp \rightarrow 
\infty)=0~.
\label{conditions}
\end{align}
As discussed in Ref. \cite{noir}, due to rotational invariance in the unpolarized case, Eq. (\ref{fteff}) reduce to:
\begin{align}
  f(k_\perp) = 2 \pi \int db_\perp ~b_\perp J_0(b_\perp 
k_\perp) \tilde f(b_\perp)~.
\end{align}
The above  can be 
expanded as follows \cite{rapid}:
\begin{align}
\label{34l}
 f(k_\perp) = \sum_{n=0}^\infty {  
k_\perp^{2n}  \langle b^{2n}_\perp \rangle 
 } P_n^{J_0}\,,
\end{align}
where the $P_n^{J_0}$ are the expansion coefficients 
of the Bessel 
function 
and $\langle b^{2n} \rangle$ are weighted moments containing  the dynamical 
information on the partonic proton structure.
Let us remind that then mean partonic distance can be defined by means of the 
probability distribution in a standard way:
\begin{align}
 \langle b^2_\perp \rangle = \int d^2 b_\perp~b_\perp^2\tilde f(b_\perp)~.
\end{align}
In the following subsections we discuss the main steps to get a lower and upper 
bounds for $\langle b^2_\perp \rangle$ given a measured $\sigma_{eff}$ once the scenario Eq. (\ref{sap}) is assumed.
\subsubsection{The minimum}
 Let us start with the minimum. By using the properties previously discussed 
(\ref{conditions}) one can show that:
\begin{align}
\int_0^\infty 
dk_\perp~ f(k_\perp)^{s-1} {d \over
dk_\perp}f(k_\perp)=-{1 \over s}~,
\label{exp2}
\end{align}
with $s > 0$. In Ref. \cite{rapid}, 
in order to evaluate the minimum of the mean transverse distance, a 
useful 
relation between the integral Eq. (\ref{sap}) and $\langle b^2_\perp
\rangle$ has been found.
To this aim let us define the following function:
\begin{align}
\label{id20}
     d_2(k_\perp) = -2 \frac{f'(k_\perp)}{k_\perp}= -4 
\sum_{n=1}k_\perp^{2n-2} \langle b^{2n} \rangle~ 
P_n^{J_0} n~.
\end{align}
For simplicity we use the notation $\langle b^2_\perp \rangle \equiv 
\langle 
b^2 \rangle$. One may notice that $\langle b^2 \rangle =d_2(k_\perp=0)$, similarly to 
the case of the standard form factors and  the charge  radius of the proton.
The above function is normalized as follows:
\begin{align}
 \int_0^\infty dk_\perp~k_\perp d_2(k_\perp)=2~.
\end{align}
By using the identity Eq. (\ref{exp2}) with $s = 3$ one gets
 \begin{align}
-\frac{1}{3} &= \int_0^\infty dk_\perp ~f(k_\perp)^2f'(k_\perp) = 
-\frac{1}{2} \int_0^\infty dk_\perp ~ k_\perp f(k_\perp)^2 
d_2(k_\perp)=&\\ \nonumber
&=
2 \int_0^\infty dk_\perp ~f(k_\perp)^2\sum_{n=1} 
k_\perp^{2n-1} \langle b^{2n} \rangle 
P_n n=\\ \nonumber
&=\int_0^\infty dk_\perp~f(k_\perp)^2 \left[-\frac{\langle b^2 
\rangle k_\perp}{2} +2  \sum_{n=2} 
k_\perp^{2n-1} \langle b^{2n} \rangle 
P_n n  \right]=\\ \nonumber
&= -\int_0^\infty dk_\perp~f(k_\perp)^2 \frac{\langle b^2 
\rangle k_\perp}{2} +2  \sum_{n=2} 
 \langle b^{2n} \rangle 
P_n n  \int_0^\infty dk_\perp~ k_\perp^{2n-1} f(k_\perp)^2~,
\end{align}
where the expansion in Eq. (\ref{id20}) has been used.  
 The above expression can be rearranged  to obtain:
\begin{align}
\label{last}
 \int_0^\infty dk_\perp~k_\perp f(k_\perp)^2 = \frac{2}{3 \langle 
b^2\rangle} +4 \sum_{n=2} P_n n \frac{\langle b^{2n}\rangle}{\langle 
b^2 \rangle} \int_0^\infty dk_\perp~ f(k_\perp)^2 k_\perp^{2n-1}~.
\end{align}
 In the above equations we set $P_n = P_n^{J_0}$ for the sake of brevity.
By using variance property,  \textsl{i.e.} $\langle b^n \rangle \geq 
\langle b \rangle^n$, one can show that the second term on the right hand side 
of Eq. (\ref{last}) is positive defined, thus leading to the condition:
\begin{align}
     \int_0^\infty dk_\perp~k_\perp f(k_\perp)^2 \geq \frac{2}{3 \langle 
b^2\rangle}~.
\end{align}
The above condition, combined with the definition in eq. (\ref{sap}), allows one to find a minimum for $\langle b^2 \rangle$, i.e.:
 \begin{align}
  \langle b^2 \rangle  \geq \ddfrac{\sigma_{eff}}{2 \pi}~.
 \end{align}

 \subsubsection{The maximum}
 Let us now discuss the procedure, 
 given a value of 
$\sigma_{eff}$, to obtain 
a maximum for $\langle b^2 \rangle$  in the approximation  of eq. (\ref{sap}).
 In this, more involved, case one should solve the following inequality
 \begin{align}
 \label{ah}
     \dfrac{2 \pi}{\sigma_{eff}} = \int_0^\infty dk_\perp~k_\perp 
     f(k_\perp)^2 \leq \dfrac{1}{N \langle b^2 \rangle}~,
 \end{align}
 with $N$ a generic real number.
The above expression is equivalent to  the following:
  \begin{align}
\int_0^\infty dk_\perp~k_\perp f(k_\perp)
\Big[ N f(k_\perp)\langle b^2 \rangle -d_2(k_\perp)  \Big] \leq 0~.     
 \end{align}
 The sufficient, but not necessary, condition to solve the above inequality is:
  \begin{eqnarray}
  N \langle b^2 \rangle f(k_\perp) \leq d_2(k_\perp)~.
 \end{eqnarray}
 By using the series expansion of  $f(k_\perp)$ and $d_2(k_\perp)$, 
 Eqs. (\ref{34l}-\ref{id20}) respectively, and by using the variance property, 
one gets:
 \begin{align}
 N  \sum_{n=0} P_n k_\perp^{2n} \langle 
b^{2n+2} \rangle \leq   \sum_{n=0} \frac{P_n}{n+1} k_\perp^{2n} \langle 
b^{2n+2} 
\rangle~.
\end{align}
 By shifting from $n$ to $n =\tilde n -1$, one then  obtains:
  \begin{align}
  N  \sum_{\tilde n=1} P_{\tilde n -1} k_\perp^{2 \tilde n -2} \langle 
b^{2 \tilde n} \rangle \leq   \sum_{\tilde n=1} \frac{P_{\tilde n -1} 
}{\tilde n} k_\perp^{2 \tilde n-2} \langle 
b^{2\tilde n} 
\rangle~.
\end{align}
In principle one can solve the above inequality by comparing equal powers 
of $k_\perp$, i.e.:  $P_{\tilde n-1} N \leq 
P_{\tilde n -1}/\tilde n$. Since the function $P_n$ changes sign with $n$, 
one finds:
\begin{align}
\label{chain1}
\left\lbrace
 \begin{array}{l}
N \leq \dfrac{1}{\tilde n}~~\mbox{for $\tilde n$ odd}\\
\\
\\
N \geq \dfrac{1}{\tilde n}~~\mbox{for $\tilde n$ even}
 \end{array}
\right.
\end{align}
Therefore $P_{\tilde n-1}$ is positive for 
$\tilde n $ odd and negative for $\tilde n$ even.
Analytically one 
finds a chain of solutions:
\begin{align}
\label{chain}
\underbrace{ \frac{1}{2}}_{\tilde n =2} \leq N \leq \underbrace{ 
 \frac{1}{1} }_{\tilde n =1};~~~ \underbrace{ \frac{1}{4}}_{\tilde n 
=4} \leq N \leq \underbrace{ 
 \frac{1}{3} }_{\tilde n =3};~~.....
\end{align}
One can generalize  the above result in the following  form:
\begin{align}
\label{chain2}
 \frac{1}{\tilde n} \leq N \leq \frac{1}{\tilde 
n-1}~,~~~~\mbox{with}~~~~ \tilde n ~\mbox{even}~,
\end{align}
or, in terms of the original $n$ ($\tilde n = n+1$):
\begin{align}
\label{chain3}
 \frac{1}{ n +1} \leq N \leq \frac{1}{\
n}~,~~~~\mbox{with}~~~~ n ~\mbox{odd}~.
\end{align}
As discussed in Ref. \cite{rapid}, in order to find a truncation on the above 
chain, some conditions on the behaviour of the EFF must be imposed even if
 the EFF, defined  through  Eq. (\ref{uno}), is essentially unknown.
To this aim, we found that a comparison between the EFF and the standard one 
could guide toward a solution of the problem.
In fact,  similarly to standard case \cite{brod}, at large $k_\perp$, \textsl{i.e.} in the pQCD domain,  dynamical correlations between partons tend to decrease. In this 
case, it reasonable to expect that the  EFF would be close to the product of standard 
form factors \cite{strikman3} whose asymptotic behaviours are $1/Q^4$ (Dirac) and $1/Q^6$ 
(Pauli). These conditions 
could  be not true in all domain of $k_\perp$ but they are expected  in the large $k_\perp$ limit, allowing to cut the chain in Eq. (\ref{chain}).
On a more quantitative level,
the condition required to solve the inequality (\ref{ah}) is that, at large 
$k_\perp$, the function 
$f(k_\perp)$ should fall to zero at least as $ k_\perp^{-2r} $ with $r> 1$.
As discussed in Ref. \cite{rapid}, 
this conjecture is supported by 
 all model calculations of dPDF 
(even those not built up to calculate dPDFs). In 
particular let us mention that one of the most used dPDF ansatz 
makes use of EFF which is the product of the gluon form factor  which
satisfies the asymptotic condition mentioned above.
Under the hypothesis that the EFF falls
off at large $k_\perp$ as $k_\perp^{-2r}$ with $r > 1$, then the $n=1$ 
contribution to the chain  (\ref{chain}) is the dominant one, thus:
\begin{eqnarray}
 \dfrac{1}{2}\leq N \leq 1~.
\end{eqnarray}
In particular, since in Eq. (\ref{ah}) we are interested in $1/N$, we found that
$ 1\leq 1/N \leq 2$.
Collecting these results one finds:
\begin{eqnarray}
 \dfrac{2 \pi}{\sigma_{eff}} = \int_0^\infty dk_\perp~k_\perp 
 f(k_\perp)^2 \leq \dfrac{2}{\langle b^2 \rangle}~,
\end{eqnarray}
which is the desidered result. Combining all results, one gets:
 \begin{align}
\label{old}
{\sigma_{eff} \over 3 \pi} \leq \langle b^2 \rangle \leq  {\sigma_{eff} 
\over \pi}~,
\end{align}
which is the main result of Ref. \cite{rapid}. 
The above relation has been checked within all models of the EFF in the 
literature.
Let us remark that in order  to make contact with experimental 
extraction of  $\sigma_{eff}$,
this result has been obtained under the approximation of Eq.(\ref{sap}).
Thanks to this feature, data on $\sigma_{eff}$ have been converted in 
the range of $\langle b^2 
\rangle$ \cite{rapid}.  
In the following Sections we will describe how $\sigma_{eff}$ can be 
generalized to 
include partonic perturbative and non perturbative correlations, thus 
breaking the factorized ansatz in Eq. (\ref{sap}), and discuss
how these correlations modify 
the relationship
 between $\langle b^2 
\rangle$ and $\sigma_{eff}$ in Eq.(\ref{old}).

\subsection{Generalization to 2v1 case}  
\begin{figure*}[t]
\centering
\hskip 1cm
\includegraphics[scale=0.6]{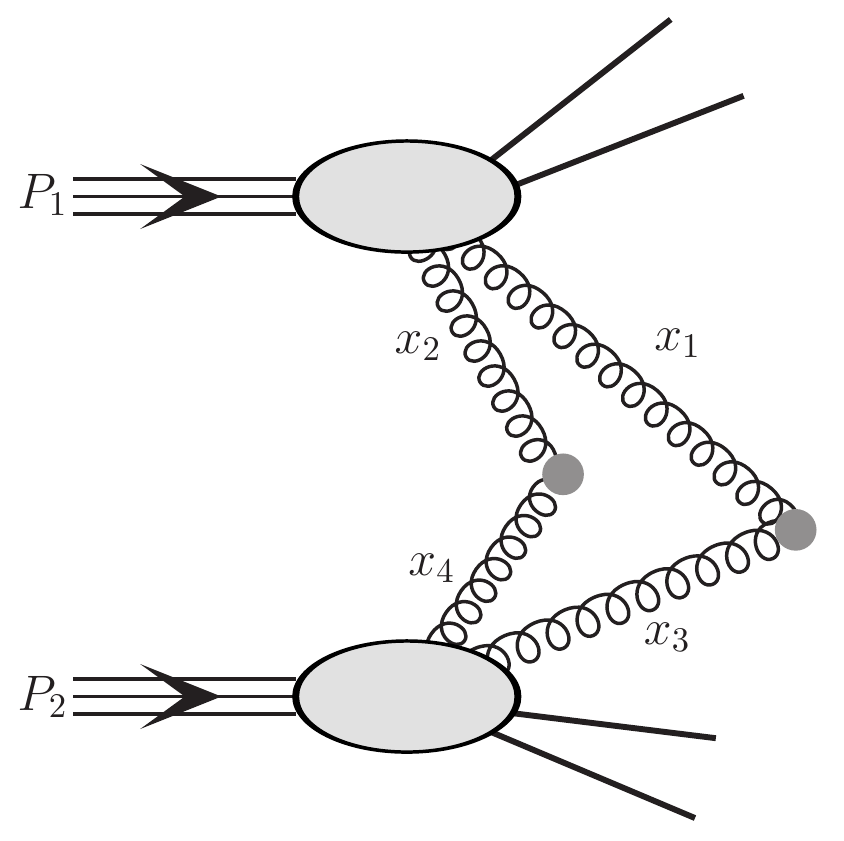}
\hskip 3cm
\includegraphics[scale=0.6]{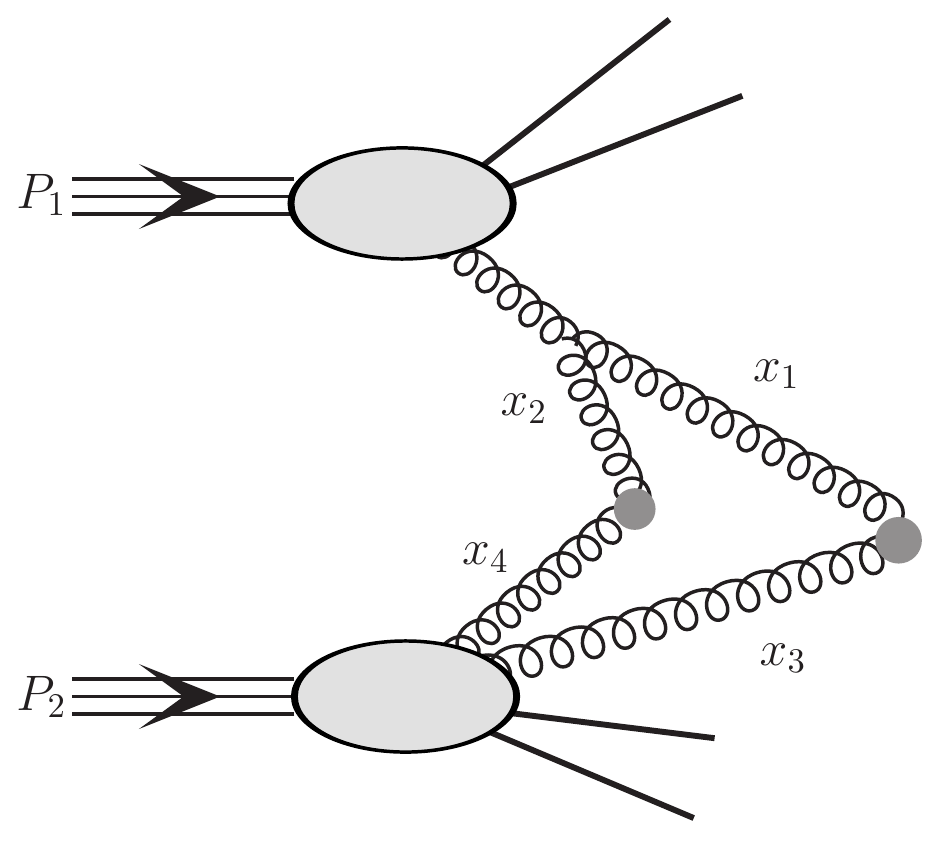}
\vskip 1cm
\caption{ \textsl{Diagrammatic representation of the two contributions 
to a DPS process: 
the so called 2v2 mechanisms is shown in the left panel and
the 2v1 mechanism in the right panel.
Small grey blobs represent the hard scattering processes.}}
\label{1v2}
\end{figure*}

As discussed in 
Ref. 
\cite{Kasemets:2012pr,ww_25_1,Gaunt:2009re,gauntladder,
Treleani:2018dbg,Ryskin:2011kk,77,Gaunt:2012dd}, the DPS cross section might 
receive a contribution from the so called 2v1 
mechanism. In this case, 
one parton pair active in the processes is 
perturbatively produced from the 
splitting of a single parton, 
\textsl{e.g.} $g\rightarrow g g$,
see the right panel of Fig. \ref{1v2}.
Given the large gluon flux at LHC energies, such a contribution can be
non-negligible \cite{77} for double quarkonia and/or Higgs production 
\cite{gauntladder} with respect to  
the standard 2v2 
mechanism shown in the left panel of Fig. \ref{1v2}. This contribution 
breaks the simple ansatz in Eq. (\ref{sap}) and it is of pure perturbative 
origin. 
Its presence in dPDF evolution equation and in DPS cross sections has been 
carefully investigated
\cite{ww_25_1,Treleani:2018dbg,Ryskin:2011kk,Diehl:2018kgr,add1,Gaunt:2012dd}.  
{Within this mechanism, the separation 
of the parton pair is set by the hard scale in the splitting, 
$1/Q \ll b \ll 1/\Lambda$. Since one typically assumes that the 
non-perturbative $b$-profile has a width of order $1/\Lambda$, one can 
approximate $b=0$ in the 2v1 term\cite{ww_3_1,gauntladder,75,Gaunt:2012dd}.}
In this Section we consider  the formalism developed in  Refs. 
\cite{gauntladder}, where the $\sigma_{eff}$ definition is generalized to
include the 2v1 contribution. As discussed in Ref. 
\cite{gauntladder}, one can decompose the total DPS cross section in terms 
of the two leading 2v2 and 2v1 
contributions  as follows
\begin{align}
 \sigma^{DPS} = \frac{\Omega^{2v2}}{\sigma_{eff,2v2} } 
+\frac{\Omega^{2v1}}{\sigma_{eff,2v1} }~,
\end{align}
where here $\Omega^{2v2}$ and $\Omega^{2v1}$
represent the DPS cross sections 
calculated with longitudinal double PDFs for both mechanisms, and 
weighted by their corresponding $\sigma_{eff}$.
In particular the $\Omega^{2v1}$ term is calculated with dPDFs whose 
initial condition is given by the 
splitting term alone at the initial scale \cite{gauntladder}. 
As discussed in Ref. \cite{gauntladder},
in experimental analyses
it is usually assumed that
$\sigma^{DPS}=\Omega^{2v2}/\sigma_{eff}$. Within this approach, one can  
incorporate the 2v1 contribution in 
$\sigma^{DPS}$ by using the
following generalization of $\sigma_{eff}$:
 \begin{align}
 \frac{1}{\sigma_{eff} } = \frac{1}{\sigma_{eff,2v2} } + 
\frac{1}{\sigma_{eff,2v1}} \frac{\Omega^{2v1} }{\Omega^{2v2} }~. 
\label{sefftot}
\end{align}
Under the assumption that the longitudinal  dependence of dPDFs 
factorizes from the transverse 
one, the effective cross sections for the two mechanisms read:
\begin{align}
\label{sef22}
 \frac{1}{\sigma_{eff,2v2}} &= \int  \frac{ d^2 k_\perp}{(2 \pi)^2} 
f(k_\perp)^2\,, 
\\
 \frac{1}{\sigma_{eff,2v1}} &= \int  \frac{ d^2 k_\perp}{(2 \pi)^2} 
f(k_\perp)= \tilde f({\bf b_\perp}=0)\,, 
\label{sef21}
\end{align}
where it is worth noticing that both the above expression depends on the 
same effective form factor,
$f(k_\perp)$, so that they  are not independent quantities. 
The first equation is the standard one, see Eq. (\ref{sap}).
The second one reflects the 
perturbative production of the couple of partons, occurring approximately at
zero relative distance in 
transverse plane.
In terms of the present notation, 
the main result of  Ref. \cite{rapid} reads:
\begin{align}
\label{di0}
   \dfrac{1}{3\pi \langle 
b^2 \rangle }  \leq  \dfrac{1}{\sigma_{eff,2v2}} \leq \dfrac{1}{\pi 
\langle 
b^2 \rangle }~,
\end{align}
where, by following Ref. \cite{gauntladder},
$\sigma_{eff,2v2}$ represents the usual definition of $\sigma_{eff}$ if 
only 
the 2v2 mechanism is considered, see Eq. (\ref{sef22}).
In the case where also the 2v1 mechanism is included in the analysis, in order
to relate
$\langle b^2 \rangle$ to the experimentally extracted $\sigma_{eff}$ Eq. 
(\ref{sefftot}),
we need first to find 
a relation between the mean partonic distance and $\sigma_{eff,2v1}$, 
defined in 
Eq. (\ref{sef21}) and appearing in the full definition of $\sigma_{eff}$ 
in Eq. 
(\ref{sefftot}).
To this aim, we  can derive a new expression of the relevant integral with a
procedure similar to the one already described in the first part of this 
section for the 2v2 
mechanism:
\begin{equation}
 \int_0^\infty dk_\perp~k_\perp f(k_\perp) 
 = \dfrac{1}{ \langle b^2 \rangle } + 
4 \sum_{n=2} \dfrac{ \langle b^{2n}\rangle  P_n^{J_0}n }{ \langle b^2 
\rangle } 
\int_0^\infty dk_\perp~k_\perp^{2n-1} f(k_\perp)~.
\end{equation}
Due to variance properties, the overall sign of the second 
term of the above equation is positive and consequently:
\begin{align}
 \dfrac{1}{\sigma_{eff,2v1}  } \geq \dfrac{1}{2\pi \langle b^2 
\rangle}~.
 \label{di1}
\end{align}
Furthermore, similarly to the 2v2 case , in order to 
estimate a reasonable maximum, one needs solve  the following inequality:
\begin{align}
 \int_0^\infty dk_\perp~k_\perp \left[ \bar N f(k_\perp)- 
\dfrac{d_2(k_\perp)}{2}    
\right] \leq 0~,
\end{align}
where $\bar N$ is an arbitrary unknown number. Under the additional 
assumption 
that the EFF falls to zero at large $k_\perp$ at least as fast as 
$k_\perp^{-2r}$ with $r \ge 1$, one finds  the desired condition: \begin{align}
  \dfrac{1}{\sigma_{eff,2v1}  } \leq \dfrac{2}{\pi \langle b^2 
\rangle}~.
  \label{di2}
 \end{align}
Linking Eq. (\ref{di1}) and Eq. (\ref{di2}), 
the following result is found:
\begin{align}
 { \sigma_{eff,2v1} \over 2 \pi} \leq \langle b^2 \rangle 
 \leq { 2~ \sigma_{eff,2v1} \over \pi }~. \label{new2v1}
\end{align}
Combining Eq.(\ref{di0}) and Eq.(\ref{new2v1}) in Eq. (\ref{sefftot}) 
one obtains the final inequality:
\begin{align}
 \label{lastt}
 \dfrac{ \sigma_{eff}}{3 \pi} \left(1+ \dfrac{3}{2}r_v  \right) \leq 
 \langle b^2\rangle \leq  \dfrac{ \sigma_{eff}}{ \pi 
} \left( 1+ 2 r_v  \right)~,
\end{align}
where here we have defined the ratio
$r_v = \Omega_{2v1} / \Omega_{2v2}$~, with $r_v \geq 0$.
Let us remark that, in principle, the ratio $r_v$ could depend on the
rapidities of particles produced in the final state and hence on parton 
fractional momenta in the 
initial state \cite{gauntladder}. Such a dependence is not explored in 
the present analysis.  
The difference between the maximum and the minimum in Eq.(\ref{lastt}) 
gives an estimate of the theoretical error
on the transverse distance of the two active partons:
\begin{align}
 \Delta = { \sigma_{eff} \over \pi } {2 \over 3} \left(1+  {9 \over 4} 
r_v 
\right)~.
\end{align}
The main effect of the inclusion of the 2v1 mechanism is to 
shift 
the $\langle b^2\rangle$ range towards higher values and 
to increase its theoretical error with respect to  the case where 
$r_v=0$. 
In particular, the  comparison between the $r_v \neq 0$ and $r_v=0$ 
cases, makes sense only if $\sigma_{eff}$ is assumed to be equal in both 
scenarios. In 
principle, as observed in Refs. \cite{ww_25_1,gauntladder}, in order to 
observe $\sigma_{eff} \sim 15$ mb, one should expect $\sigma_{eff,2v2} 
\sim 30$ mb.
In general, if $r_v \neq$1, from Eq. (\ref{sefftot}) one gets 
$\sigma_{eff} \leq \sigma_{eff,2v2}$. 

We find interesting to check the validity of Eq. 
(\ref{lastt}) by using two phenomenological models for EFF, such as 
those 
described in Refs. \cite{ww_25_1,gauntladder,strikman3}.
The first one is  Gaussian EFF of the type:
\begin{align}
f(k_\perp)= e^{-k_\perp^2 a}~.
\label{fk_gaussian}
\end{align}
In this case the mean partonic distance can be obtained in term of the 
width parameter $a$ as:
\begin{align}
\langle b^2 \rangle = - 
2 \frac{d}{k_\perp d k_\perp} f(k_\perp)  \Bigg |_{k_\perp=0} = 4a~,
\end{align}
so that, according to Eqs. (\ref{sef22},\ref{sef21}),
\begin{equation}
\sigma_{eff,2v2} = 2 \langle b^2 \rangle \pi~,
\;\;\;
\sigma_{eff,2v1} = \langle b^2 \rangle  \pi~.
\end{equation}
By using the above expressions in Eq. (\ref{sefftot}), one gets the 
following 
result:
\begin{align}
\langle b^2 \rangle = \frac{ \sigma_{eff} }{ \pi } 
\left( \frac{1}{2} +r_v  \right)~,
\end{align}
which is included in the range Eq. (\ref{lastt}).
As a second example we
consider an EFF which 
is the square of the gluon form factor \cite{strikman3}, i.e.:
\begin{align}
f(k_\perp) = \left( \frac{k_\perp^2}{m_g^2}+1  \right)^{-4}~,
\label{fk_strikman}
\end{align}
with the parameter $m_g$ has been fixed by fitting HERA data, i.e.  
$m_g^2 \sim 1.1$ GeV$^2$ \cite{strikman3}.
In this case one obtains:
\begin{align}
\langle b^2 \rangle = - 2 \frac{d}{k_\perp d k_\perp} f(k_\perp) 
\Bigg |_{k_\perp=0} =  {16 \over m_g^2} ~,
\end{align}
and, according to Eqs. (\ref{sef22},\ref{sef21}),
\begin{equation}
\sigma_{eff,2v2} = {7 \over 4} \langle b^2 \rangle \pi~,~
\sigma_{eff,2v1} = {3 \over 4} \langle b^2 \rangle\pi~.
\end{equation}
By using the above expressions in Eq. (\ref{sefftot}), one gets the 
following result:
\begin{align}
\langle b^2 \rangle = \frac{ \sigma_{eff} }{ \pi } \left( \frac{4}{7} + 
\frac{4}{3} r_v  \right)~,
\end{align}
which again lies in the range indicated in Eq. (\ref{lastt}).
\begin{figure*}
\centering
\includegraphics[scale=0.6]{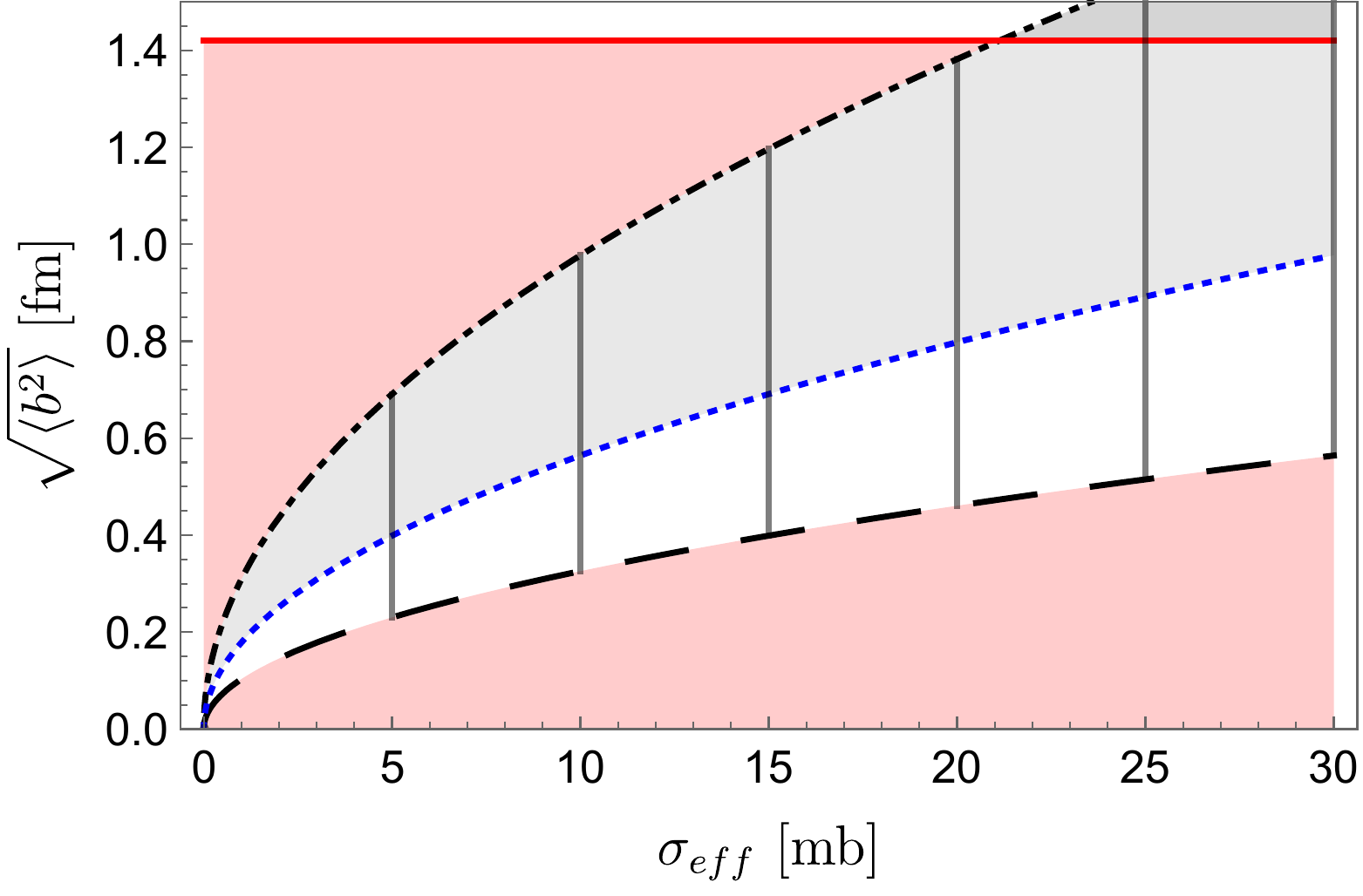}
\vskip 0.5cm
\caption{ \textsl{ 
The averaged partonic interdistance as a function of $\sigma_{eff}$. 
Dashed lines represent the minimum of Eq. (\ref{goal4}). Dot-dashed 
lines stand for the maximum of Eq. (\ref{goal4}). Dotted lines are for 
the maximum of Eq. (\ref{old}). 
Vertical lines represent the full range of allowed partonic distances 
between two partons if the 2v2 and the 2v1 mechanism effects on the 
total $\sigma_{eff}$ are not disentangled.
The shadow between lines represents the additional theoretical error 
w.r.t. the case where only the  2v2 mechanism is considered, \ie Eq. 
(\ref{old}). The areas outside lines represent the exclusion region of 
the allowed transverse distance between two partons active in a DPS 
process. The red line stands for twice the transverse proton radius.}}
\label{ind}
\end{figure*}
\noindent
Such a generalization of the inequality in Eq. (\ref{old})  is 
however process dependent, in fact, as discussed in Refs. 
\cite{ww_25_1,gauntladder}, $r_v$ 
is related to the kinematic conditions and to the type of the considered 
DPS process. Without a 
precise knowledge on $r_v$, a determination of the range of the allowed 
mean partonic transverse distance is therefore prevented.  
Nonetheless, we note  that 
$r_v \geq 0$ since it is a ratio of cross sections which is positive definite.
{Furthermore one expects that  $r_v \leq 1$, at least for processes involving 
small parton fractional momenta, as those typically considered in the present 
analysis.
Additionally, in this regime, 
one expects the 2v1 mechanism to be subdominant in the 
pQCD evolution of dPDFs w.r.t. the 2v2 one, being the former 
proportional to the gluon density and the latter proportional to its 
square}. Thanks to these features, 
for $r_v=0$ one finds the  minimum in Eq. (\ref{lastt}) while 
for  $r_v=1$  one  finds its maximum:
\begin{align}
 \dfrac{ \sigma_{eff}}{3 \pi} \leq 
 \langle b^2\rangle \leq  \dfrac{3 \sigma_{eff}}{ \pi
}~.
    \label{goal4}
\end{align}
This result allows one to obtain information on the interpartonic 
distance of two active partons in a DPS process without knowing details 
on the relative size of the two mechanisms, 2v1 and 2v2,  \textsl{i.e.} 
the exact knowledge of  $r_v$. This, of course, comes at the expense of 
an increased theoretical error. In order to quantify such an effect, we 
have plotted in Fig. \ref{ind} the two extremes of Eq. (\ref{goal4}) 
with dashed and dot-dashed lines respectively,    together with the 
maximum of Eq. (\ref{lastt}) evaluated with $r_v=0$ (dotted lines), as 
function of different values of $\sigma_{eff}$. 
The white area between the curves represents the theoretical error 
associated to the 2v2 mechanism alone. The shaded area 
represents the additional  uncertainty induced by the particular choice 
on $r_v$ leading to Eq. (\ref{goal4}). 
In addition, in Table \ref{tabnew}, we report 
the interpartonic distances, calculated according to Eq. (\ref{lastt}), 
 for $\sigma_{eff}$ values extracted from a selection of experimental 
analyses. 
It should be noted that, in all cases,  $\sqrt{\langle b^2 \rangle} < 2 
R_\perp=1.42$ fm, where $R_\perp \sim 0.71$ fm is the transverse 
electro-magnetic proton radius. 
\begin{table}[t]
\begin{center}
\centering
\begin{tabular}{|c|c|c|c|c|}
\hline
Ref. & Process  &$\sigma_{eff}$  $\mbox{[mb]}$ &   $ \sqrt{{\sigma_{eff} \over 3 \pi 
}}$ [fm] & $\sqrt{{ 3 \sigma_{eff} \over  \pi }}$[fm] \\ \hline
\cite{data13} &D0$~(J/\Psi+\Upsilon),~\sqrt{s}=1.96$ TeV  & $\sim 2.2$  
& 0.15 & 0.46 \\
\cite{data15} & $(J/\Psi+Z),~\sqrt{s}=8$ TeV& $\sim 4.7$ & 0.22 & 0.67  
\\
\cite{data14} &D0$~(J/\Psi+J/\Psi),~\sqrt{s}=1.96$ TeV & $\sim 4.8$ & 
0.23 & 0.68  \\
\cite{data16} &$(W+J/\Psi),~\sqrt{s}=7$ TeV & $\sim 6.1$ & 0.25 & 0.76  
\\
\cite{data11} &ATLAS $(J/\Psi+J/\Psi),~\sqrt{s}=8$ TeV & $\sim 6.3$ & 
0.26 & 0.78  \\
\cite{data12} &$(J/\Psi+J/\Psi),~\sqrt{s}=7$ TeV & $\sim 8.2$  & 0.29 & 
0.88 \\
\cite{data8} &LHCb $(J/\Psi+J/\Psi),~\sqrt{s}=13$ TeV & $\sim 8.8$  & 
0.31 & 0.92 \\
\cite{data6} &ATLAS (4-jets) $\sqrt{s}=7$ TeV & $\sim 14.9$  & 0.40 & 
1.19  \\
\cite{data17} &LHCb $(\Upsilon+c\bar{c}),~\sqrt{s}=7-8$ TeV & $\sim 18.0$  & 0.44 
& 1.31 \\
\cite{data9} &CMS (W+2-jets)  $\sqrt{s}=7$ TeV & $\sim 20.7$  & 0.47 & 
1.41 \\ \hline
\end{tabular}
\caption{\textsl{Ranges of mean transverse distance evaluated by means 
of Eq. (\ref{goal4}) sorted by increasing values of $\sigma_{eff}$ as extracted from the quoted experimental analyses.}}
\label{tabnew}
\end{center}
\end{table}
We close this Section by observing that  Eq. (\ref{lastt}) can be 
inverted to give:
\begin{align}
    {\pi \langle b^2 \rangle \over (1+2 r_v)} \leq \sigma_{eff} \leq   
{\pi \langle b^2 \rangle 3 \over  \left( 1 + {3 \over 2 }r_v \right )    
 }~.
    \label{inverse}
\end{align}
In such a form, given the value
of $\langle b^2 \rangle$ 
associated to a particular $f(k_\perp)$, the inequality in Eq. (\ref{inverse}) 
predicts 
the expected range in 
$\sigma_{eff}$ associated to that specific model.
Most importantly, Eq.(\ref{inverse}) shows that, given an EFF, 
characterized by 
$\langle b^2 \rangle$,
the $\sigma_{eff}$ value 
does depend on the relative size of the 2v1 contribution. In particular,
if $r_v$ is significantly larger than zero, 
the corresponding $\sigma_{eff}$ will be lower than the one obtained if 
the 2v2 mechanism alone were considered ($r_v=0$).

\subsection{Generalization to the unfactorized ansatz}
\noindent
As shown in several constituent quark model  calculations of dPDFs,
double parton correlations 
may survive at high
momentum {transfer, which are relevant for experimentally measurable processes
\cite{Mel_19,noiold,noi1,noir,noij2,noiww}}. 
In this Section we investigate how Eq. (\ref{old}) is 
generalized to the case in which the factorized ansatz  is not assumed 
thus allowing the presence of longitudinal 
and mixed longitudinal-transverse partonic correlations.
We consider an unapproximated scenario in which 
$\sigma_{eff}$ depends on the longitudinal momentum fractions of the 
active partons, as suggested in Refs. \cite{noiplb1,noiads}.
Within this improved framework, the relationship  between $\sigma_{eff}$ 
and the mean partonic distance will be sensitive to $x_1-x_2$ 
correlations. 
For this purpose we consider the simplest generalization of the results presented in Section \ref{proton}, namely we consider non-factorizable dPDFs,
in the zero rapidity case, \textsl{i.e.} $x_i =x_i'$. 
{For processes
whose production is dominated by gluons, as those discussed in this paper, the expression 
for  $\sigma_{eff}$ can be simplified to \cite{noiplb1}:} 
\begin{align}
 \sigma_{eff}(x_1,x_2)=2\pi \ddfrac{ \big[ F(x_1)F(x_2) \big]^2 }{\int
 dk_\perp~k_\perp F(x_1,x_2,k_\perp)^2  }~,
  \label{ssg}
\end{align}
{where $F(x)$ and 
$F(x_1,x_2)$ represent 
single and double gluon PDFs, respectively.}    
We  assume
that  the $k_\perp$ dependence of 
dPDFs has the same behaviour
and asymptotics as the ones discussed for the EFF in Sec. \ref{proton}.
 This feature is inspired by the GPDs behavior, whose dependence on  
the transverse momentum basically follows the one of the related form 
factor.
In the present case the inequality is obtained following the same steps 
outlined in Section 4, but retaining the full
$(x_1,x_2)$ dependence via dPDFs.
To this aim we expand the form factor as:
\begin{align}
 F(x_1,x_2,k_\perp) &= 2 \pi \int db_\perp~b_\perp J_0(k_\perp b_\perp) 
\tilde F(x_1,x_2,b_\perp)=
\\
\nonumber
&=  \sum_{n=0}^\infty k_\perp^{2n}~ P_n^{J_0} \int d^2 b_\perp 
~b_\perp^{2n} \tilde F(x_1,x_2,b_\perp)~.
\end{align}
By using the definition in 
Eq. (\ref{dx4}) one can derive the following expression:
\begin{align}
\label{gen2}
  \dfrac{ F(x_1,x_2,k_\perp)}{F(x_1,x_2,0)} = \sum_{n=0}^\infty {  
k_\perp^{2n}  \langle b_\perp^{2n} \rangle_{x_1,x_2} 
 } P_n^{J_0}\,.
\end{align}
\subsubsection{The minimum}
The strategy to get a relation between $\sigma_{eff}$ and $\langle b_\perp^2\rangle_{x_1,x_2}$ is 
then very similar to the one 
we discussed in Sect. 4.1.1. 
Eqs. (\ref{exp2}, \ref{id20})
can be generalized to
\begin{align}
\int_0^\infty 
dk_\perp~ F(x_1,x_2,k_\perp)^{s-1} {d \over 
dk_\perp}F(x_1,x_2,k_\perp)=-{F(x_1,x_2,0)^s \over s}~,
\label{id2}
\end{align}
and the derivative function to:
\begin{align}
d_2^{x_1,x_2}(k_\perp) = -\ddfrac{2}{F(x_1,x_2,0)} 
\ddfrac{F'(x_1,x_2,k_\perp)}{k_\perp} = -4 \sum_{n=1}^\infty k_\perp^{2n-2} 
\langle \bp^{2n} \rangle_{x_1,x_2} P_n n~. 
\end{align}
In this case, $\langle \bp^2 \rangle_{x_1,x_2}=d_2^{x_1,x_2}(k_\perp=0)$.
Within these settings we generalize Eq. (\ref{last}) to obtain 
the first relation between $\sigma_{eff}(x_1,x_2)$ and $\langle \bp^2 
\rangle_{x_1,x_2}$. As in Sect. 4.1.1, we start with Eq. (\ref{id2}) for $s=3$:
\begin{align}
 -\ddfrac{F(x_1,x_2,0)^3}{3} &= \int_0^\infty dk_\perp~F(x_1,x_2,k_\perp)^2 
F'(x_1,x_2,k_\perp)
\\
\nonumber
&=  -\ddfrac{F(x_1,x_2,0)}{2}\int_0^\infty dk_\perp~k_\perp 
F(x_1,x_2,k_\perp)^2 d_2^{x_1,x_2}(k_\perp)
\\
\nonumber
&= F(x_1,x_2,0) 2 \int_0^\infty dk_\perp~F(x_1,x_2,k_\perp)^2 \sum_{n=1}^\infty 
k_\perp^{2n-1} \langle \bp^{2n} \rangle_{x_1,x_2} P_n n
\\
\nonumber
&=F(x_1,x_2,0) \int_0^\infty dk_\perp~F(x_1,x_2,k_\perp)^2 \left[- 
\ddfrac{\langle \bp^2 \rangle_{x_1,x_2}k_\perp }{2}+2 \sum_{n=2}^\infty 
k_\perp^{2n-1} 
\langle \bp^{2n} \rangle_{x_1,x_2} P_n n \right]
\\
\nonumber
&=-F(x_1,x_2,0)\int_0^\infty dk_\perp~k_\perp F(x_1,x_2,k_\perp)^2 
\ddfrac{\langle \bp^2 \rangle_{x_1,x_2} }{2} 
\\
\nonumber
&+ 2F(x_1,x_2,0) 
\sum_{n=2}^\infty  \langle \bp^{2n} \rangle_{x_1,x_2} P_n n \int_0^\infty 
dk_\perp~k_\perp^{2n-1}F(x_1,x_2,k_\perp)^2~.
\end{align}
Finally one gets:
\begin{align}
 \int_0^\infty dk_\perp~k_\perp F(x_1,x_2,k_\perp)^2 &= \dfrac{2}{3} 
\ddfrac{F(x_1,x_2,0)^2}{ \langle \bp^2 \rangle_{x_1,x_2}}
\\
\nonumber
&+4 \sum_{n=2}^\infty  \ddfrac{ \langle \bp^{2n} \rangle_{x_1,x_2}}{ \langle 
\bp^{2} \rangle_{x_1,x_2}} \int_0^\infty dk_\perp~F(x_1,x_2,k_\perp)^2 
k_\perp^{2n-1}~.
\end{align}
By noticing that the second term 
is positive definite, one obtains
the following inequality:
\begin{align}
 \int_0^\infty dk_\perp~k_\perp F(x_1,x_2,k_\perp)^2 \geq \dfrac{2}{3} 
\ddfrac{F(x_1,x_2,0)^2}{ \langle \bp^2 \rangle_{x_1,x_2}}~.
\end{align}
In terms of $\sigma_{eff}(x_1,x_2)$
defined in Eq. (\ref{ssg}) the result is recast into:
\begin{align}
\label{min12}
  \ddfrac{ \langle \bp^2 \rangle_{x_1,x_2}}{r_{gg}(x_1,x_2)^2 }  \geq 
\ddfrac{\sigma_{eff}(x_1,x_2) }{3 \pi}~, 
\end{align}
where we defined:
\begin{align}
 r_{gg}(x_1,x_2) = {   F(x_1,x_2,k_\perp=0) \over F(x_1)F(x_2) 
}~.
\label{rggg}
\end{align}

\subsubsection{The maximum}
\begin{figure*}[h]
\centering
\includegraphics[scale=0.50]{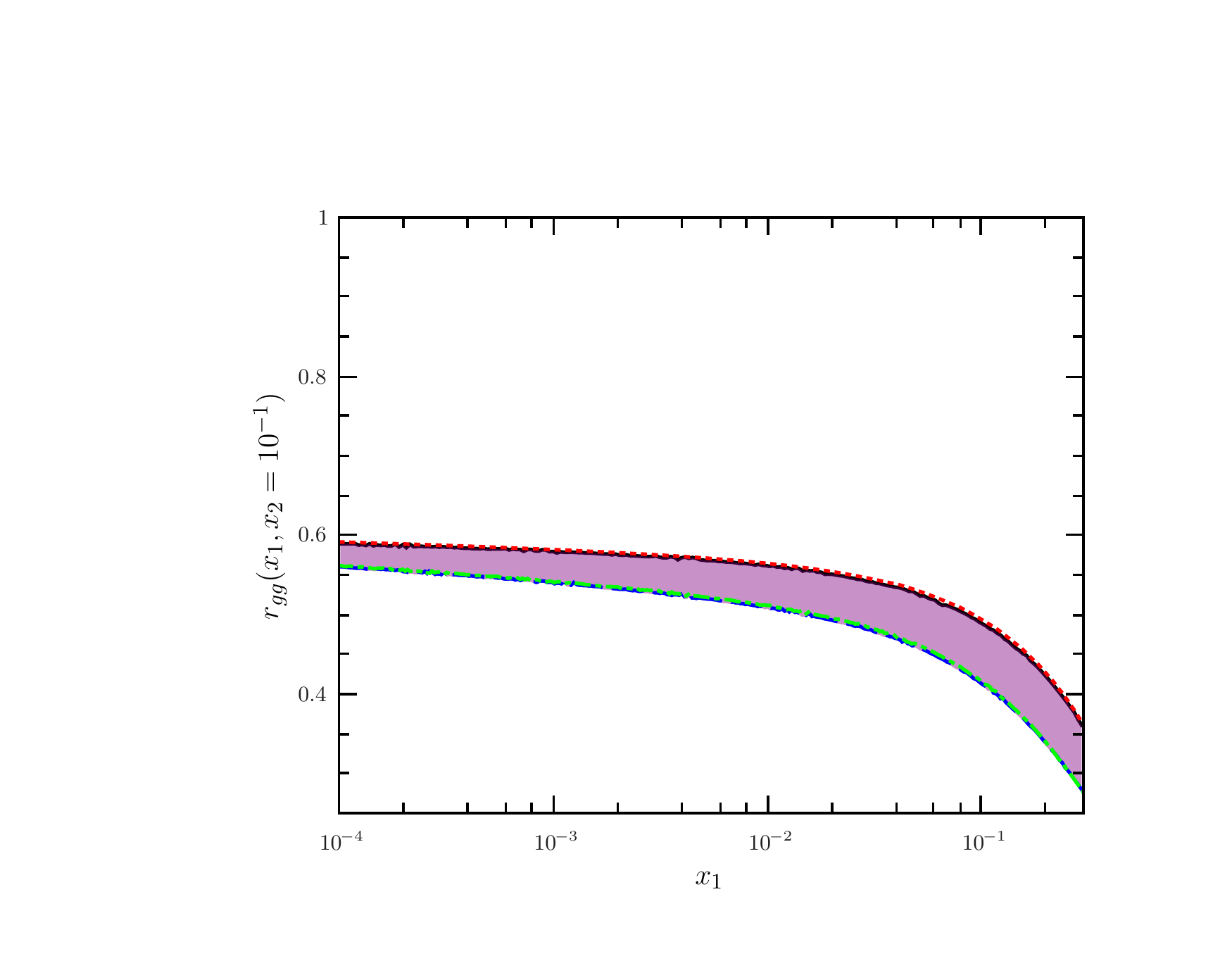}
\hskip -2.cm
\includegraphics[scale=0.50]{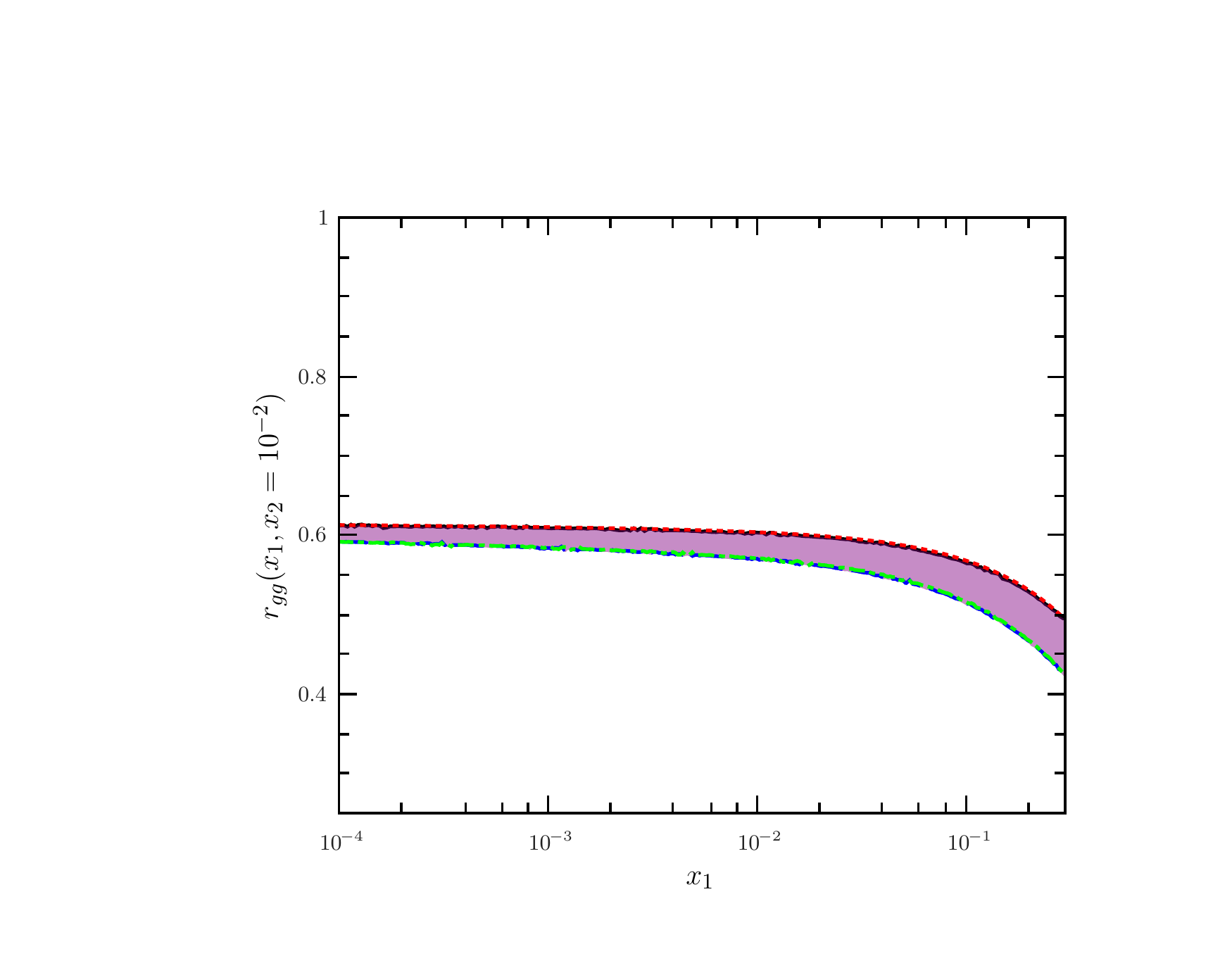}\\ 
\vskip -2cm
\includegraphics[scale=0.50]{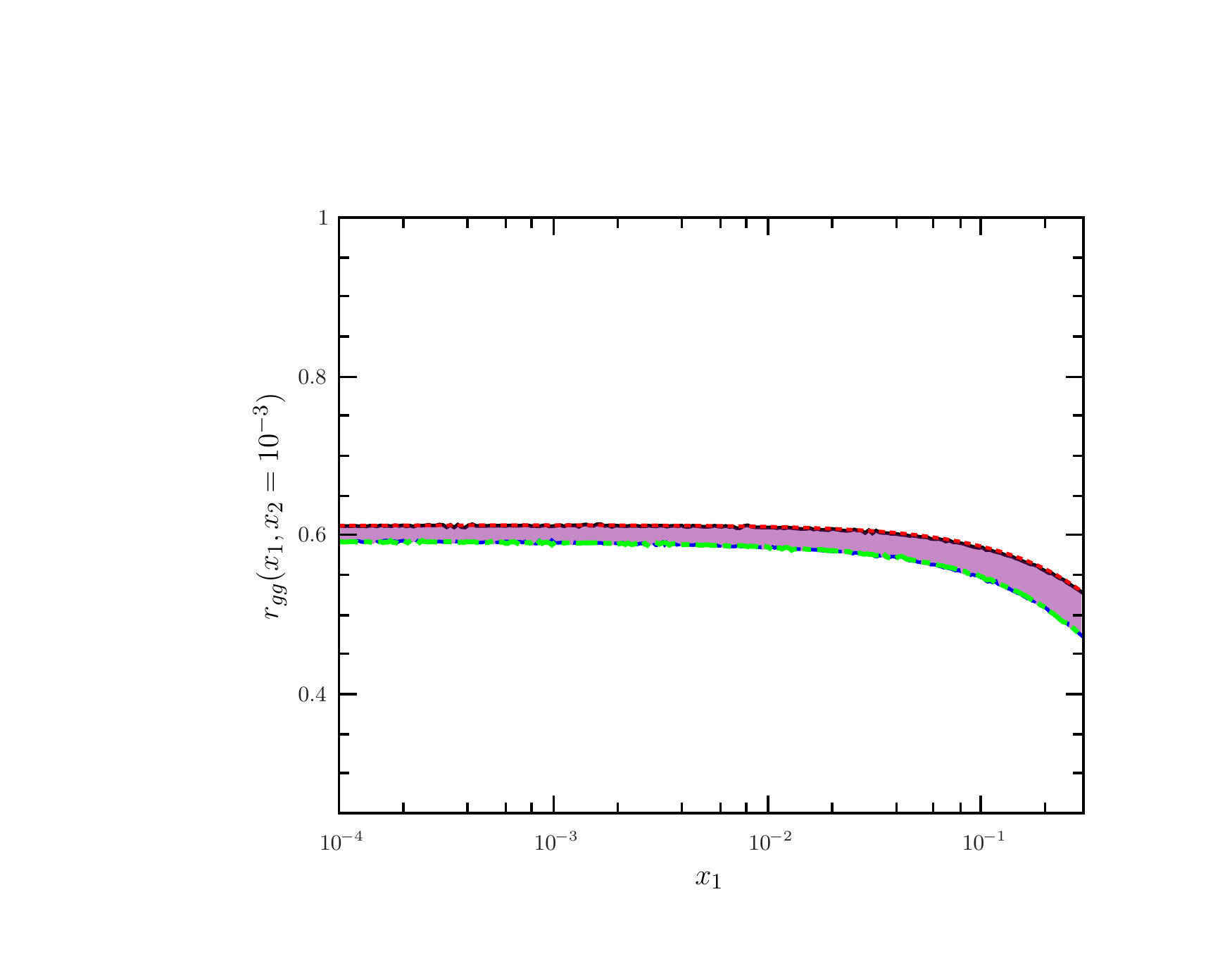}
\hskip -2cm
\includegraphics[scale=0.50]{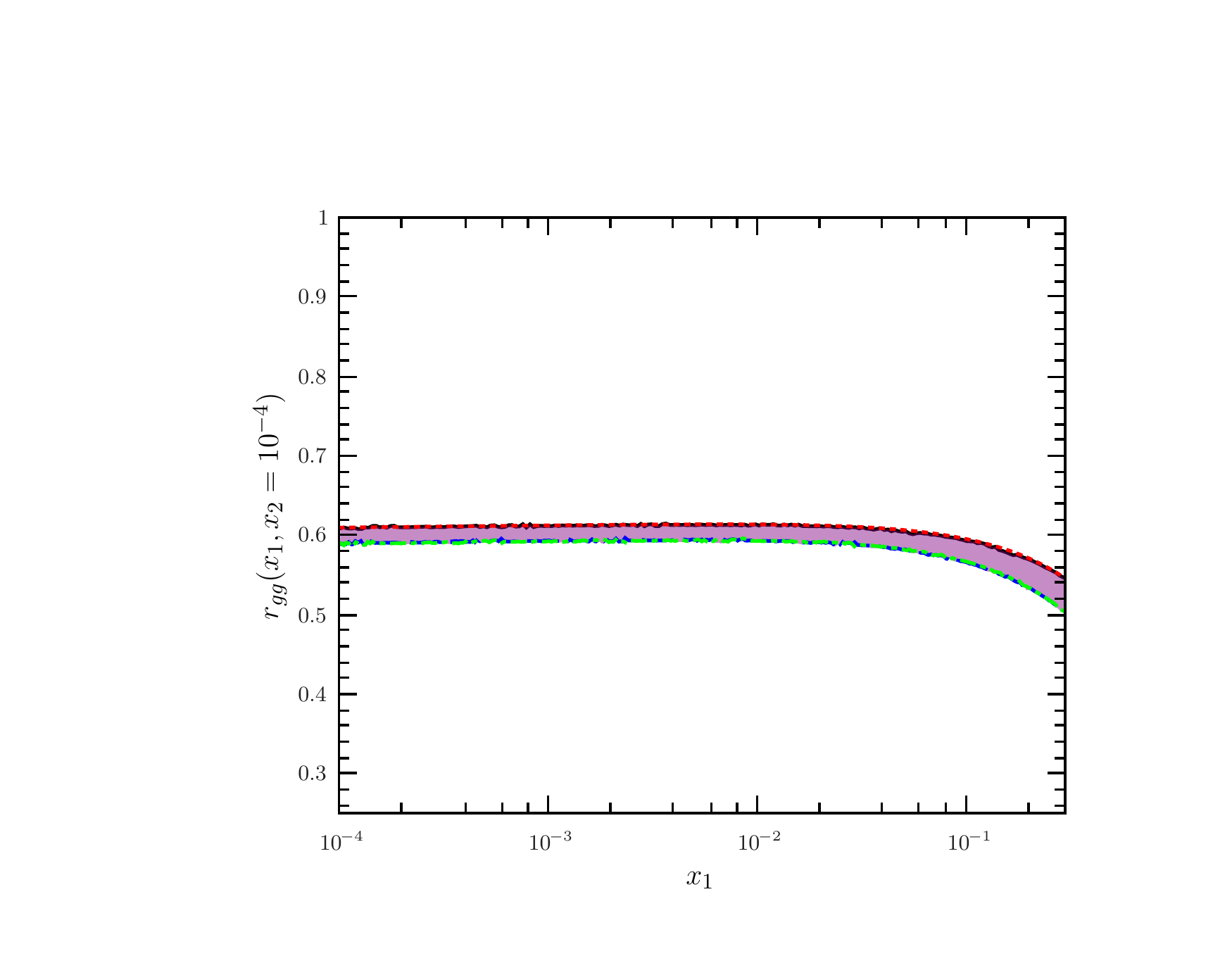}
\vskip -0.5cm
\caption{ \textsl{The ratio $r_{gg}(x_1,x_2)$, Eq. (\ref{rggg}), 
evaluated 
at $x_2=10^{-1}$ (upper-left panel), $x_2=10^{-2}$ (upper-right panel),
$x_2=10^{-3}$ (lower-left panel) and $x_2=10^{-4}$ (lower-right panel). 
The HO model predictions are indicated 
by full line ($Q^2=4m_c^2$)
and dashed line ($Q^2=m_H^2$).
The HP model predictions are indicated by dotted line ($Q^2=4m_c^2$) 
and dot-dashed line ($Q^2=m_H^2$). The band stands for the difference 
between the 
calculations performed in the two scales.}} 
\label{f_rgg}
\end{figure*}

In this last part we derive the maximum of the mean transverse distance for a unfactorized 
dPDFs ansatz. 
Also in this case we consider a general realistic 
condition, \textsl{i.e.}  the $k_\perp$ dependence of dPDFs is dominated by that of the 
EFF. In this scenario, the only difference between this case and the previous 
one, see Sec. \ref{ss1}, is that for fixed values of $x_1$, $x_2$ and the 
energy scale, the dPDF at $k_\perp=0$ can be different from 1. In this case it is necessary to find a value of $N$ such that:
\begin{align}
 \label{12eq1}
 \ddfrac{2\pi}{\sigma_{eff}(x_1,x_2)} = \ddfrac{ \int_0^\infty dk_\perp~k_\perp F(x_1,x_2,k_\perp)^2  }{\Big[F(x_1) F(x_2) \Big]^2 } \leq \ddfrac{1}{N 
\langle  \bp^2 \rangle_{x_1,x_2} }~.
\end{align}
The above expression can be rearranged as follows:
\begin{align}
\label{12eq2}
   \int_0^\infty dk_\perp~k_\perp 
F(x_1,x_2,k_\perp) \left[F(x_1,x_2,k_\perp)  N 
\ddfrac{r_{gg}(x_1,x_2)^2 }{F(x_1,x_2,0)^2 }-1    \right] \leq 0~.
\end{align}
In order to find a sufficient condition to solve the above inequality, we make 
use of the following identity:
\begin{align}
 \int_0^\infty dk_\perp ~k_\perp F(x_1,x_2,k_\perp) d_2^{x_1,x_2}(k_\perp) 
=F(x_1,x_2,0)~.
\end{align}
By using the above expression, Eq. (\ref{12eq2}) becomes:
\begin{align}
 \int_0^\infty dk_\perp~k_\perp \ddfrac{ F(x_1,x_2,k_\perp) 
}{F(x_1,x_2,0)}  \left[  \ddfrac{F(x_1,x_2,k_\perp)  }{ 
F(x_1,x_2,0) } N r_{gg}(x_1,x_2)^2 \langle \bp^2 
\rangle_{x_1,x_2}-d_2^{x_1,x_2}(k_\perp)    \right] \leq 0~,
\end{align}
and a sufficient condition to solve the inequality reads:
\begin{align}
 \ddfrac{F(x_1,x_2,k_\perp)  }{ 
F(x_1,x_2,0) } N r_{gg}(x_1,x_2)^2 \langle \bp^2 
\rangle_{x_1,x_2} \leq d_2^{x_1,x_2}(k_\perp)~. 
\end{align}
By using Eqs. (\ref{gen2}-\ref{id2}) the latter can be rewritten as:
\begin{align}
 N(x_1,x_2) \sum_{\tilde n=1} P_{\tilde n-1}k_\perp^{2 \tilde n-2} 
\langle \bp^{2 \tilde n} \rangle_{x_1,x_2} \leq \sum_{\tilde n=1} 
\ddfrac{P_{\tilde n-1}}{\tilde n}  k_\perp^{2 \tilde n-2} 
\langle \bp^{2 \tilde n} \rangle_{x_1,x_2}~,
\end{align}
where we define $N(x_1,x_2) \equiv N r_{gg}(x_1,x_2)^2$. The same chain of solutions shown in Eqs. (\ref{chain1}-\ref{chain3}) is obtained, the main difference being now that these solutions correspond to $N(x_1,x_2)$. Therefore one gets:
\begin{align}
 1 \leq \dfrac{1}{N(x_1,x_2) } \leq 2~,
\end{align}
which corresponds to:
\begin{align}
 r_{gg}(x_1,x_2)^2 \leq \dfrac{1}{N} \leq 2 ~r_{gg}(x_1,x_2)^2~.
\end{align}
By using this relation in Eq. (\ref{12eq1}) one finds:
\begin{align}
\label{max12}
 \ddfrac{\langle \bp^2 \rangle_{x_1,x_2} }{r_{gg}(x_1,x_2)^2 } \leq \ddfrac{\si 
}{\pi}
\end{align}
Combining Eq. (\ref{min12}) and Eq. (\ref{max12}) one finally obtains:
\begin{align}
 {\sigma_{eff}(x_1,x_2)  \over 3 \pi} 
  \leq   \dfrac{\langle b^2 \rangle_{x_1,x_2}}{r_{gg}(x_1,x_2)^2} \leq
  {\sigma_{eff}(x_1,x_2)  \over  \pi}~, 
 \label{goal3}
\end{align}
which, with respect to Eq. (\ref{old}),  additionally depends on the ratio $r_{gg}$.
Such a ratio  encodes longitudinal correlations 
in the proton structure,
and therefore so does  $\langle b^2 \rangle_{x_1,x_2}$.
\begin{table}[t]
\begin{center}
\centering
\begin{tabular}{c|c|c|c|c|c|c|c|c|c|c|c|} 
\hline              
\multicolumn{3}{|c|}{Kinematics}&\multicolumn{4}{c|}{HO 
model}&\multicolumn{4}{c|}{HP 
model} \\ 
\cline{4-11}
\multicolumn{3}{|c|}{}&\multicolumn{2}{c|}{ 
${\sigma_{eff}(x_1,x_2)\over 3 \pi}$ }&\multicolumn{1}{c|}{$ {\langle 
b^2 
\rangle_{x_1,x_2}\over r_{gg}^2(x_1,x_2)}$}&\multicolumn{1}{c|}{ 
${\sigma_{eff}(x_1,x_2)\over \pi}$}
&\multicolumn{2}{c|}{$  {\sigma_{eff}(x_1,x_2)\over 3 \pi}$}&
\multicolumn{1}{c|}{${\langle b^2 \rangle_{x_1,x_2}\over 
r_{gg}^2(x_1,x_2)}$}&\multicolumn{1}{c|}{${\sigma_{eff}(x_1,x_2)\over 
\pi}$} \\ 
\hline
\multicolumn{3}{|c|}{$x_1=10^{-4},x_2=10^{-4}$  
} &  \multicolumn{2}{c|}{0.263} &  \multicolumn{1}{c|}{0.429}   & 
\multicolumn{1}{c|}{0.790}&  \multicolumn{2}{c|}{0.235}   & 
\multicolumn{1}{c|}{0.425} & 
\multicolumn{1}{c|}{0.704} \\ 
\multicolumn{3}{|c|}{ $x_1= 10^{-2},x_2=10^{-4}$ } 
&  \multicolumn{2}{c|}{0.256}  &  \multicolumn{1}{c|}{0.405} & 
\multicolumn{1}{c|}{0.767}&  \multicolumn{2}{c|}{0.227}   & 
\multicolumn{1}{c|}{0.462}&  \multicolumn{1}{c|}{0.680}  \\ 
\multicolumn{3}{|c|}{$x_1=10^{-2},x_2=10^{-2}$ } 
&  \multicolumn{2}{c|}{0.268} &  \multicolumn{1}{c|}{0.370}  & 
\multicolumn{1}{c|}{0.805}&  \multicolumn{2}{c|}{0.226}   & 
\multicolumn{1}{c|}{0.453}&  \multicolumn{1}{c|}{0.678} \\
\hline
\end{tabular}
\caption{ \textsl{Numerical test of the validity of Eq. (\ref{goal3}) 
by using the HO and HP models. Double gluon distributions have been 
evolved at 
$Q^2= m_H^2$. The various entries in the table are expressed in 
[$\mbox{fm}^2$].
}}
\label{table1}
\end{center}
\end{table}  
In order to test the inequality,
we have  evaluated the terms appearing in Eq. (\ref{goal3}), \textsl{i.e.} 
$\sigma_{eff}(x_1,x_2)$, $r_{gg}(x_1,x_2)$ and 
$\langle b^2 \rangle_{x_1,x_2}$, 
by using  quark model calculations of dPDFs and PDFs,  see Fig. \ref{f_rgg} for 
  $r_{gg}(x_1,x_2)$.
Since we are interested in kinematic regions close to those  experimentally 
accessed, we have calculated the above quantities by using  
the digluon dPDF 
obtained through pQCD evolution at high momentum scales, $Q^2=m_H^2$, 
and test Eq. (\ref{goal3}) in three couples of fractional momenta, 
$x_1 = x_2 = 10^{-4}$, $x_1 = 10^{-4}, 
x_2=10^{-2}$ and  $x_1 = x_2 = 10^{-2}$. 
In addition, in order to assess the hadronic model dependence of the 
results, Eq. (\ref{goal3}) has been calculated with digluon distribution 
obtained within two different CQMs. The results are reported Table 
\ref{table1} and, as one may notice,
the inequality Eq. (\ref{goal3}) is verified in all kinematic 
conditions. 
One should also notice that, at variance with the case where the 
factorization ansatz in Eq. (\ref{ans}) is assumed,  in this new scenario the 
effects of correlations in dPDFs, embodied in the $r_{gg}(x_1,x_2)$ factor, play 
a crucial role in verifying the identity.
This generalized inequality effectively allows one to estimate the 
impact of double parton correlations on the range of allowed parton 
transverse distances.

\vskip 4cm

\subsection{The full relation}
In this final part we collect all previous results to obtain a full 
relation between $\si$ and $\langle \bp^2 \rangle_{x_1,x_2}$ in 
the zero rapidity region including also the splitting contribution.
In this case the full demonstration consists in a combination of the 
previous ones.
From Sec. 4.1,  we discussed the following system of relations if the 
splitting contribution is included:

\begin{align}
 \left\{    
 \begin{array}{l}
\dfrac{1}{\sigma_{eff}} =\dfrac{1}{\sigma_{eff,2v2} 
}+\dfrac{r_v}{\sigma_{eff,2v1} } \\
\\
 \dfrac{\sigma_{eff,2v2}  }{3\pi  } \leq \langle 
b^2\rangle \leq \dfrac{\sigma_{eff,2v2}  
}{\pi  } \\
\\
\dfrac{\sigma_{eff,2v1}  }{2\pi  } \leq \langle 
b^2\rangle \leq \dfrac{2\sigma_{eff,2v1}  
}{\pi  } 
 \end{array}
\right.
\end{align}
As shown in Sec. 4.2,
on the other hand side, if the $x_1-x_2$ and $k_\perp$ correlations are 
not neglected: 

\begin{align}
 \dfrac{\sigma_{eff,2v2}(x_1,x_2)r_{gg}^{2v2}(x_1,x_2)^2  }{3\pi  } 
\leq 
\langle 
b^2\rangle_{x_1,x_2} \leq \dfrac{\sigma_{eff,2v2}(x_1,x_2) 
r_{gg}^{2v2}(x_1,x_2)^2   
}{\pi  }
\end{align}
where by definition $r_{gg}^{2v2}(x_1,x_2) = r_{gg}(x_1,x_2)=
F(x_1,x_2,0;Q^2)/[F(x_1;Q^2)F(x_2;Q^2)]$. Let us remark here that 
within this notation, $F(x_1,x_2,0;Q^2)$ is the radiative digluon PDF as obtained from 
homogeneous evolution.
In order to include correlations between $x_1-x_2$ and $k_\perp$ also 
in the case where the splitting contribution in the pQCD evolution is 
included, one may introduce a new ratio:
\begin{align}
 r_{gg}^{2v1}(x_1,x_2) = 
  \dfrac{ F^{splitting}_{gg}(x_1,x_2,0;Q^2)}{F(x_1;Q^2)F(x_2;Q^2)}~,
\end{align}
where $ F^{splitting}_{gg}(x_1,x_2,0;Q^2)$ is the pure splitting  
contribution to digluon dPDF 
where the non dPDFs are evolved 
with inhomogeneous evolution 
equations with the non peturbative
digluon distribution set to zero 
at the intial scale. By performing the same steps previously discussed, one obtains
\begin{align}
 \dfrac{\sigma_{eff,2v1}(x_1,x_2)r_{gg}^{2v1}(x_1,x_2)^2  }{2\pi  } 
\leq 
\langle 
b^2\rangle_{x_1,x_2} \leq \dfrac{ 2 \sigma_{eff,2v1}(x_1,x_2) 
r_{gg}^{2v1}(x_1,x_2)^2   
}{\pi  }
\end{align}
Combining all terms one gets:
\begin{align}
\label{last11}
 \dfrac{\sigma_{eff}(x_1,x_2)}{3\pi}\left(r^{2v2}_{gg}(x_1,x_2)^2+ 
\dfrac{3}{2} 
r^{2v1}_{gg}(x_1,x_2)^2 r_v   \right) \leq \langle b^2 
\rangle_{x_1,x_2} \leq 
\dfrac{\sigma_{eff}(x_1,x_2)}{\pi}\Big(r^{2v2}_{gg}(x_1,x_2)^2+ 2 
r^{2v1}_{gg}(x_1,x_2)^2 r_v   \Big)~. 
\end{align}
This last expression represents the most general inequality  between the mean 
transverse partonic distance and $\si(x_1,x_2)$.
In order to avoid to model $r_v$,
we may consider the maximum range
by setting  $r_v=0$ and 
$r_v=1$ in the minimum and maximum bounds respectively:
\begin{align}
\label{last1}
 \dfrac{\sigma_{eff}(x_1,x_2)}{3\pi}\Big[r_{gg}^{2v2}(x_1,x_2)^2 \Big] \leq 
\langle 
b^2 \rangle_{x_1,x_2} \leq 
\dfrac{\sigma_{eff}(x_1,x_2)}{\pi}\Big[r_{gg}^{2v2}(x_1,x_2)^2+ 2 
r_{gg}^{2v1}(x_1,x_2)^2    \Big]~. 
\end{align}
In Ref. \cite{gauntladder}, authors introduced a ratio between the splitting 
contribution term to dPDFs versus the dPDF evolved only with the homogeneous one: 
\begin{align}
r_{ga}(x_1,x_2) = \dfrac{F_{gg}^{splitting}(x_1,x_2,0;Q^2)  
}{F_{gg}(x_1,x_2,0;Q^2)} \geq 0~.
\end{align}
 In their analysis, by considering different models and kinematic conditions, 
authors of Ref. \cite{gauntladder} estimated that $r_{ga}(x_1,x_2) \leq 0.2$.
 One should notice that this quantity  appear in Eqs. (\ref{last11},\ref{last1}) by rewriting $r_{gg}^{2v1}(x_1,x_2)$ as follows:
  \begin{align}
  r^{2v1}_{gg}(x_1,x_2) &= 
\dfrac{F_{gg}^{splitting}(x_1,x_2,0;Q^2)}{F_g(x_1;Q^2)F_{g}(x_2;Q^2)}
\\
\nonumber
&= 
\dfrac{F_{gg}^{splitting}(x_1,x_2,0;Q^2)}{F_{gg}(x_1,x_2,0;Q^2)}
\dfrac{F_{
gg}(x_1,x_2,0;Q^2)}  {F_g(x_1;Q^2)F_{g}(x_2;Q^2)}
\\
\nonumber
&= r_{ga}(x_1,x_2) \cdot r^{2v2}_{gg}(x_1,x_2)~.
 \end{align}
With this notation Eq. (\ref{last1}) becomes:
 \begin{align}
 \label{srga}
 \dfrac{\sigma_{eff}(x_1,x_2)}{3\pi}\Big[r_{gg}^{2v2}(x_1,x_2)^2 \Big] \leq 
\langle 
b^2 \rangle_{x_1,x_2} \leq 
\dfrac{\sigma_{eff}(x_1,x_2)}{\pi} r_{gg}^{2v2}(x_1,x_2)^2  \Big[1+ 2 
~r_{ga}(x_1,x_2)^2 
  \Big]~. 
\end{align}
To date there are no published data on 
$\si$ with an explicit evaluation 
of its dependence on $x_1$ and $x_2$. Therefore we can 
give an estimate of Eq. (\ref{srga}) with a costant $\si$. 
\begin{table}[t]
\begin{center}
\centering
\begin{tabular}{|c|c|c|c|c|}
\hline
Ref. & Process  &$\sigma_{eff}$  $\mbox{[mb]}$ &   $ \sqrt{{\sigma_{eff} \over 3 
\pi 
}}0.6$ [fm] & $\sqrt{{ 0.4 \sigma_{eff} \over  \pi }}$[fm] \\ \hline
\cite{data13} &D0$~(J/\Psi+\Upsilon),~\sqrt{s}=1.96$ TeV  & $\sim 2.2$  
& 0.09 & 0.17 \\
\cite{data14} &D0$~(J/\Psi+J/\Psi),~\sqrt{s}=1.96$ TeV & $\sim 4.8$ & 
0.13 & 0.25  \\
\cite{data11} &ATLAS $(J/\Psi+J/\Psi),~\sqrt{s}=8$ TeV & $\sim 6.3$ & 
0.16 & 0.28  \\
\cite{data12} &$(J/\Psi+J/\Psi),~\sqrt{s}=7$ TeV & $\sim 8.2$  & 0.18 & 
0.32 \\
\cite{data8} &LHCb $(J/\Psi+J/\Psi),~\sqrt{s}=13$ TeV & $\sim 8.8$  & 
0.18 & 0.33 \\
\cite{data17} &LHCb $(\Upsilon+ c\bar{c}),~\sqrt{s}=7-8$ TeV & $\sim 18.0$  & 0.26 
& 0.48 \\
 \hline
\end{tabular}
\caption{\textsl{Ranges of mean transverse distance evaluated
from Eq. (\ref{srga}) setting $r_{ga} \sim 0.2$, $r_{gg} \sim 0.6$ and by pretending 
that the experimental $\si$ is extracted by the non factorized dPDF.  
}}
\label{tabnew2}
\end{center}
\end{table}
In Tab. \ref{tabnew2} 
we report numerical estimates of the allowed range of $\langle b_\perp \rangle_{x_1,x_2}$, obtained  by using Eq. (\ref{srga}) in the 
worst scenario, \textsl{i.e.} $r_{ga} 
\sim 0.2$ and $r_{gg}(x_1,x_2)\sim 0.6$. 
We note that in this last inequality the theoretical errors are reduced and the range of mean 
distance is shifted towards smaller values with respect to the 
ranges reported in  Tab. 
\ref{tabnew} for the simple factorized case studied in Section 3.
We also remark that the physical information accessible relies upon  the approximations with which $\si$ is extracted.

\section[Relativistic effects in dPDFs ]{Relativistic effects in 
dPDFs }
\label{mel}
In this Section we consider 
relativistic effects on dPDFs, 
already addressed in Ref. \cite{noir}, 
and study their relevance when propagated at high momentum transfer in  
typical LHC kinematics, 
with a special emphasis on the digluon distribution.
Relativistic effects, in fact,
induce model independent correlations between $x_1-x_2$ and $k_\perp$ 
on dPDFs~\cite{noir}. Their study therefore is relevant since these kind of correlations are almost unknown, at variance with those between $x_1$ and $x_2$ 
for which there are indications from pQCD evolution and dPDF sum rules. 
Within this context, 
relativistc effects are embodied via Light-Front boosts which are
kinematical operators. The associated  Light-Front wave function is then frame 
independent and encodes additional kinematical correlations between $x$ and 
$k_\perp$ induced by these kinematical operators. 
Among the three forms of relativistic dynamics~\cite{dirac}, the Light-Front (LF) one 
has the maximum number of kinematical generators, such as LF 
boosts~\cite{dirac}.
This feature makes the LF approach suitable to implement special 
relativity for strongly interacting systems \cite{LF1,LF2,brod} and 
therefore it has been extensively used to evaluate other kind of parton 
distributions~\cite{Boffi1, boffi3,boffitmd, traini14}. We 
consider the dPDFs expression
presented in Ref. \cite{noi1}, \textsl{i.e.}:
\begin{align}
 \label{dpdf}
F_{ij}(x_1,x_2, \vec{k}_\perp) &\propto \int d\vec k_1 d\vec k_2~\Psi 
\left(\vec k_1+ {\vec k_\perp}, \vec k_2
\right)  \tau_i 
\tau_j \Psi^\dagger \left(\vec 
k_1, \vec k_2+ {\vec k_\perp}  \right) \delta  \left(x_1- 
\dfrac{k_1^+}{M_0} \right) \delta 
\left(x_1- \dfrac{k_1^+}{M_0} \right) \\
\nonumber
&\times \langle S \otimes F | \hat D_1^\dagger \hat D_1  \hat 
D_2^\dagger \hat 
D_2 
|S \otimes F 
\rangle~,
\end{align}
where $\vec k_i$ is the intrinsic three-momentum of the $i$ parton 
whose 
flavor is determined by $\tau_i$, 
$k_\perp$ is the relative transverse momentum  unbalance in the parton 
pair,  
$\Psi$ is the proton canonical (instant form) wave function in momentum 
space 
and $|S \otimes F\rangle $ is a generic 
spin-flavor state.
$M_0$ is the proton mass with constituent quarks treated as free 
particles and whose dependence on $x_i$ and $\vec k_{i \perp}$ is given 
by:
\begin{align}
 M_0^2 = \sum_{i=1}^{3} \frac{m_i^2+\vec k_{i\perp}^2 }{x_i}~,
\end{align}
being $m_i$ and $x_i$ the constituent quark mass  and longitudinal 
momentum 
fraction carried by the $i$ quark, respectively.
Here, as in Ref.~\cite{noi1}, we consider for simplicity  a factorized 
dependence between the spin-flavor and the spatial part of the proton 
wave function. 
For the sake of completeness, let us point out that the above condition can
be broken by, \eg spin-orbit effects, see Ref. \cite{noiold}. However, as will 
be discussed later on, since we focus on model independent features of dPDFs, we 
consider ratios that minimize these effects. To this aim, the HO is particularly 
suitable since, by construction, such contributions are neglected.
Thanks to the LF approach, momentum conservation is 
preserved, \textsl{i.e.} dPDFs vanish in the unphysical region  
$x_1+x_2>1$. 
The canonical proton  wave function appearing in Eq. 
(\ref{dpdf}) 
can be calculated within constituent quark models,
see e.g. Refs. 
\cite{noiold,noi1}. 
Nevertheless, the price for the use of the canonical proton wave 
function is the inclusion of boosts from the Light-Front centre of mass 
frame to the instant form one, \textsl{i.e.}{} the so called Melosh 
operators~\cite{melosh}, which  appear in the second line of 
Eq.~(\ref{dpdf})
and are defined as 
\begin{eqnarray}
 \hat D_i = \dfrac{m_i+x_iM_0 +i(k_{i x} \sigma_y - k_{iy} \sigma_x    ) 
}{\sqrt{(m_i+x_iM_0)^2+ k_{ix}^2+k_{iy}^2}}~,
\label{melo}
\end{eqnarray}
where  $\sigma_x$ and $\sigma_y$ are Pauli sigma matrices.
In particular, the Melosh operators allow 
to rotate Light-Front spin into the canonical one. 
We emphasize that for unpolarized PDFs, for which the initial proton
state is equal to the final one in the light-cone correlator,
the product of Meloshs reduce to the unity, $\hat 
D^\dagger \hat D = \mathtt{1}$. However, as shown in Ref. \cite{noi1}, 
in the 
case of dPDFs, for which in general $k_\perp \neq0$, Melosh operators 
contribute also in the case of unpolarized partons. In the present 
analysis, we are interested in $(x_i-k_\perp)$ correlations induced by 
Melosh operators on dPDFs.
\begin{figure*}[t]
\includegraphics[scale=0.52]{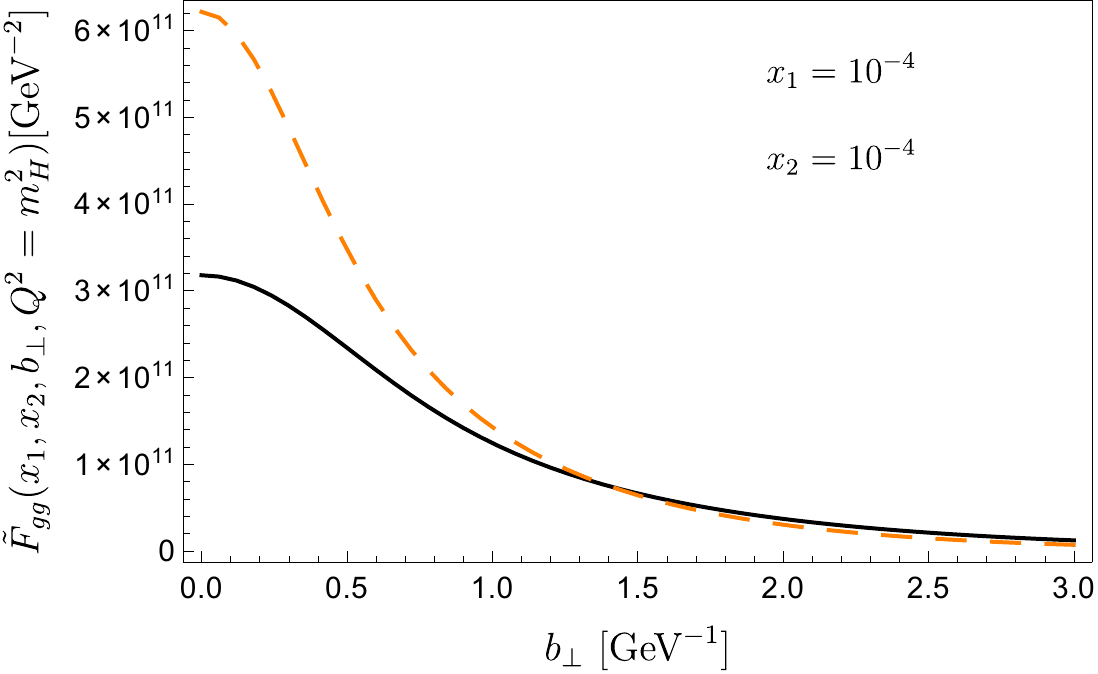}
\includegraphics[scale=0.52]{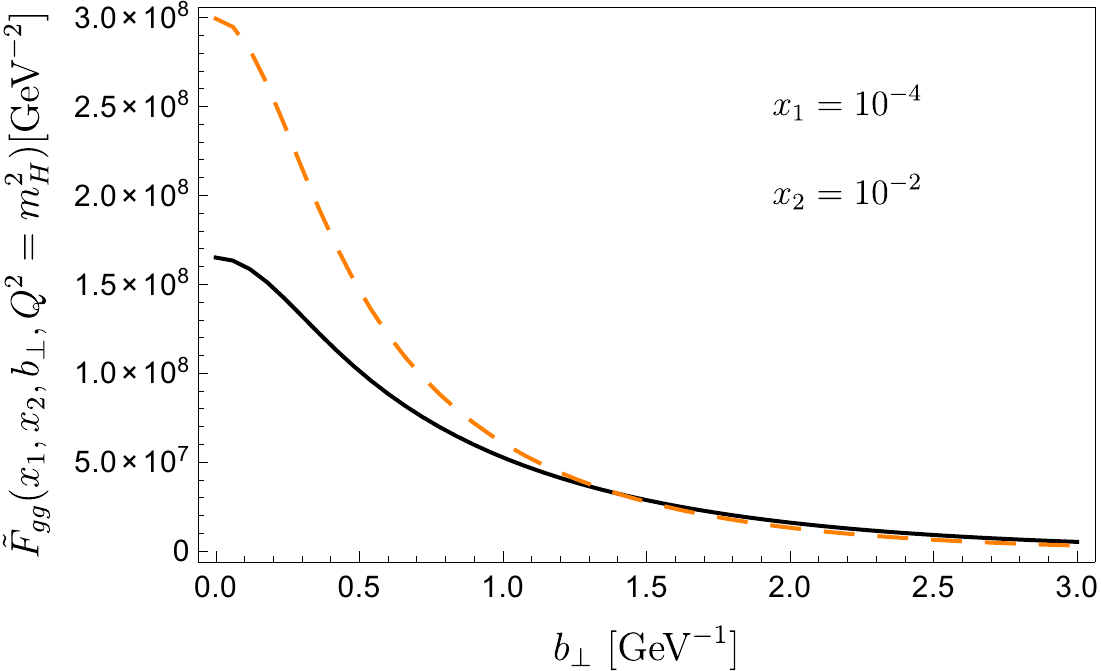}
\includegraphics[scale=0.52]
{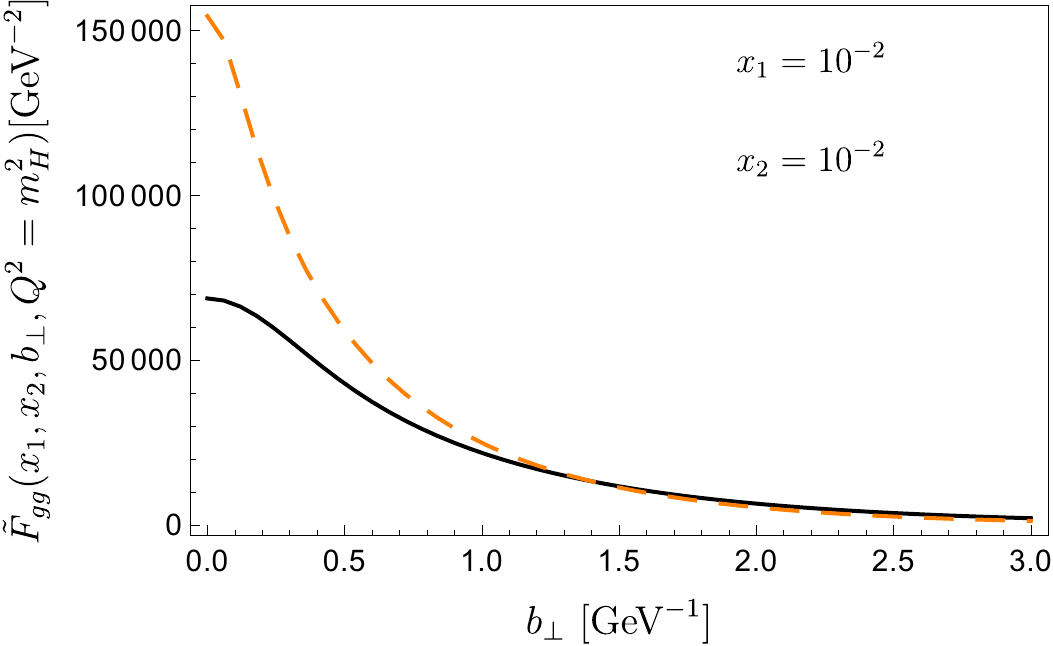}
\\
\includegraphics[scale=0.52]{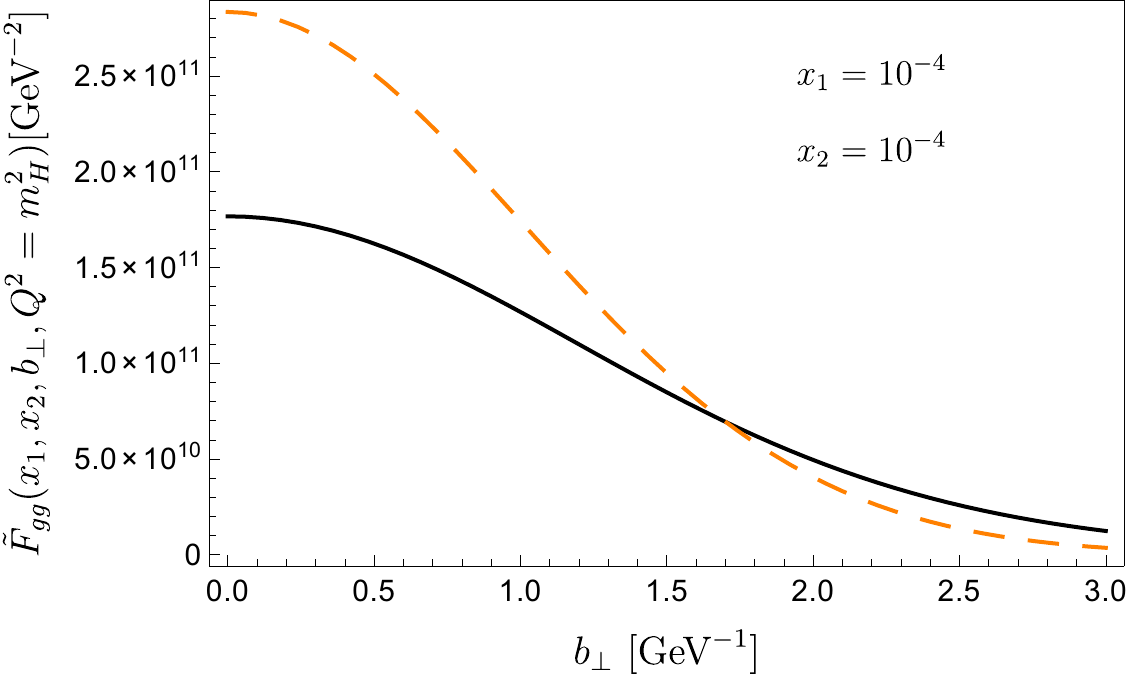}
\includegraphics[scale=0.52]{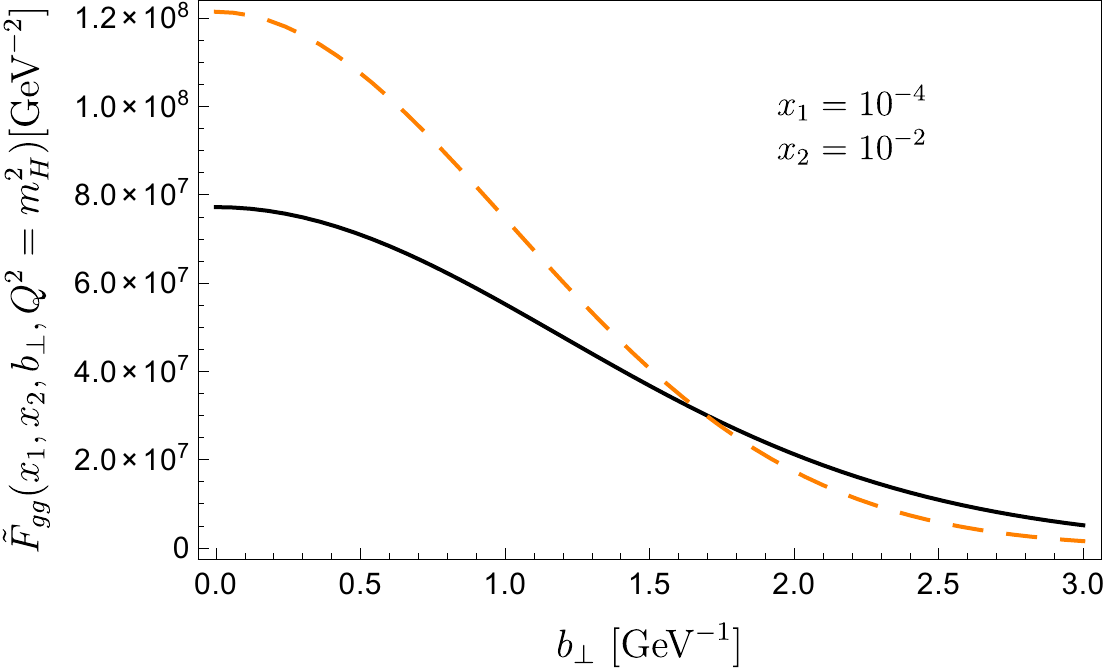}
\includegraphics[scale=0.52]{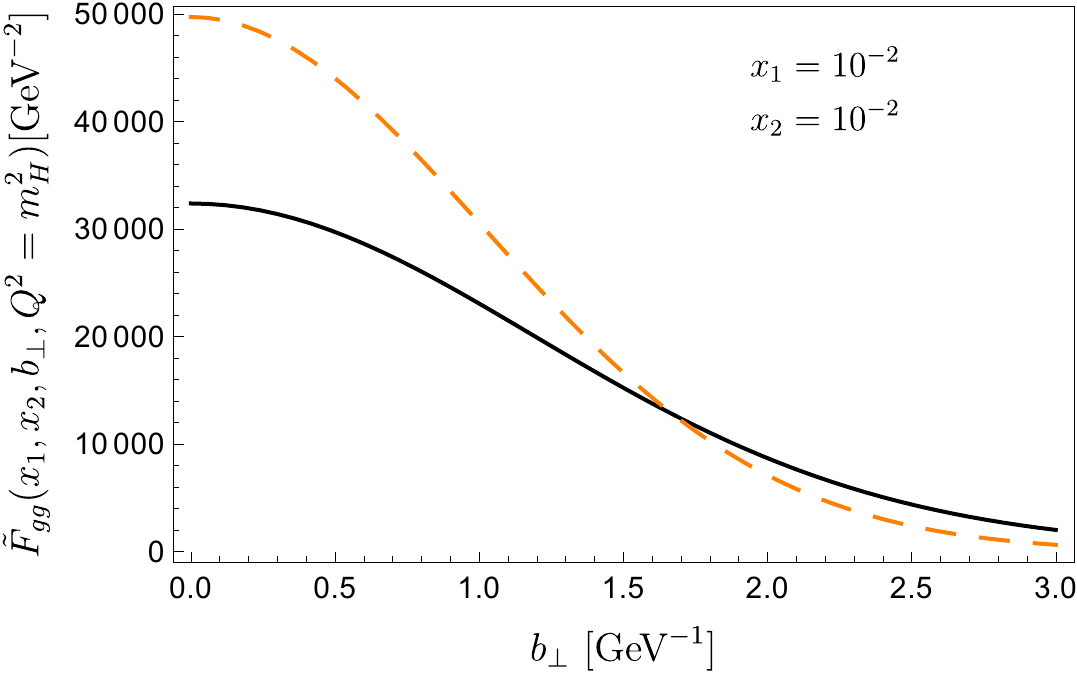}
\caption{ \textsl{The distribution $\tilde 
F_{gg}(x_1,x_2,b_\perp,Q^2=m_H^2)$ evaluated for three pairs of $x_1$ 
and $x_2$ and depending on $b_\perp$. 
Double PDF calculations are displayed with (full lines) and  without 
(dashed lines) Melosh operators. Upper panels for the HP model and lower 
panel for the HO one. }}
\label{tomo1}
\end{figure*}
However, given the complicated structure of Eq. (\ref{dpdf}), it is non 
trivial to single out their effects, since they mix with the proton wave 
function. 
In order to determine to which extent their effects on dPDFs are 
independent of the chosen hadronic model, we compare dPDF calculations 
performed within the HO and the HP  models and  build appropriate ratios 
in order to highlight relativistic effects alone.

\begin{figure*}
\includegraphics[scale=0.60]{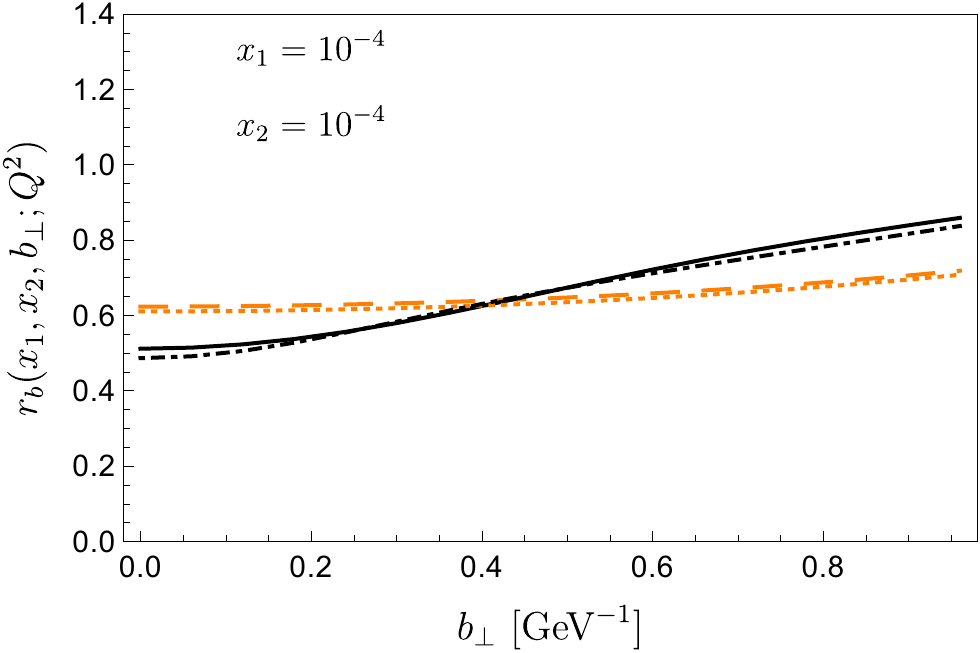}
\includegraphics[scale=0.63]{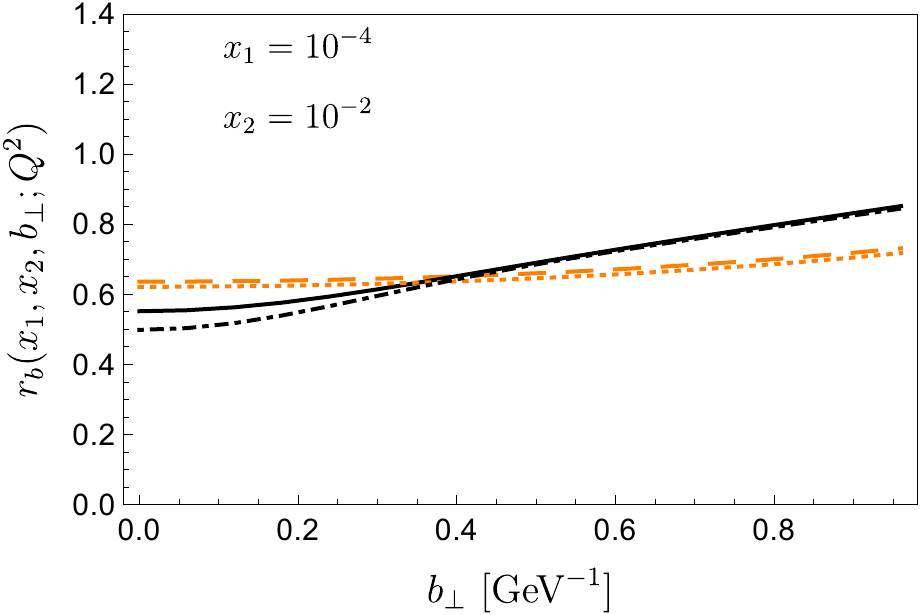}
\includegraphics[scale=0.63]{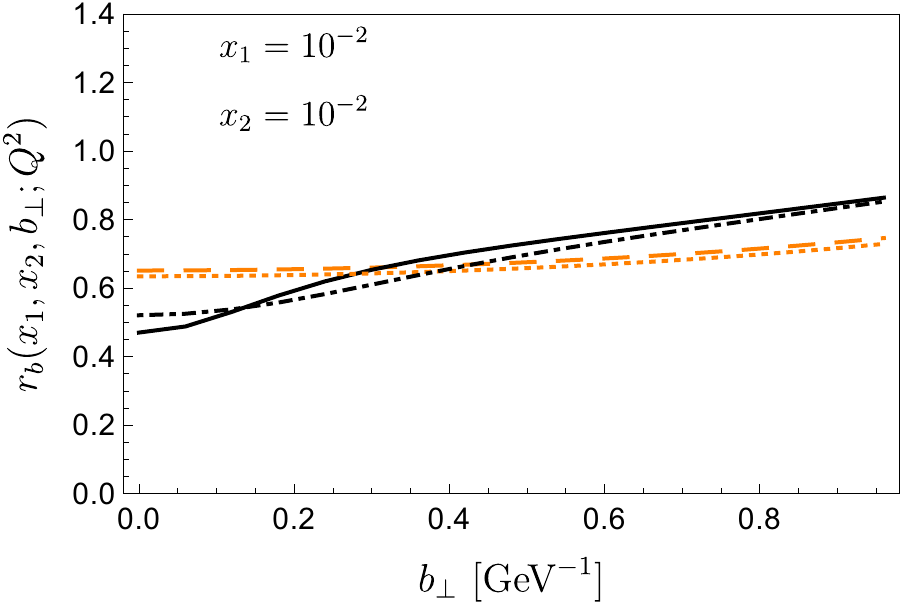}
\vskip 0.3cm
\caption{ \textsl{The ratio in Eq. (\ref{r_rb}) evaluated for digluon 
distribution  as a function of $b_\perp$ for three pairs of values of 
$x_1$ and $x_2$. 
Black lines stand for dPDF calculations within the HP model, at $Q^2 
=m_H^2$ (full) and 
at $Q^2 =4 m_c^2$ (dot-dashed).
Orange lines stand for dPDF computed within the HO model, at $Q^2 
=m_H^2$ (dashed) and 
at $Q^2 =4 m_c^2$ (dotted).}}
\label{tomo2}
\end{figure*}
\begin{figure*}[h]
\centering
\includegraphics[scale=0.6]{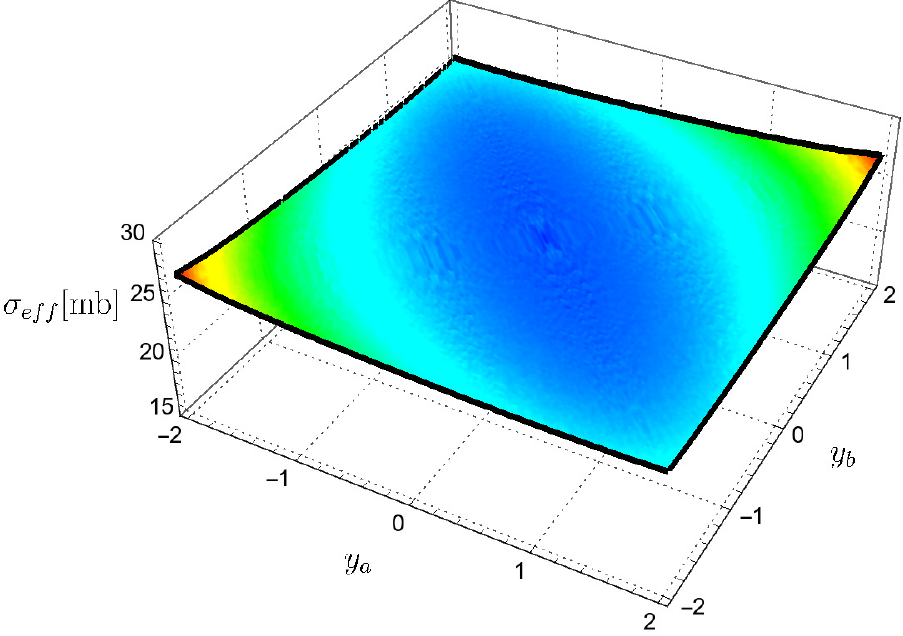}
\hskip 1cm
\includegraphics[scale=0.6]{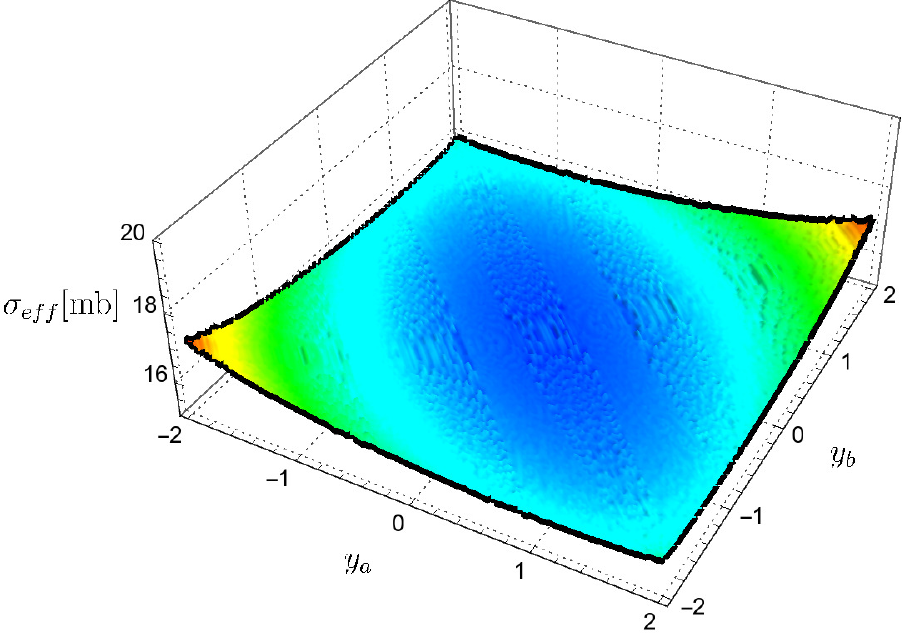}
\vskip 0.1cm
\includegraphics[scale=0.6]{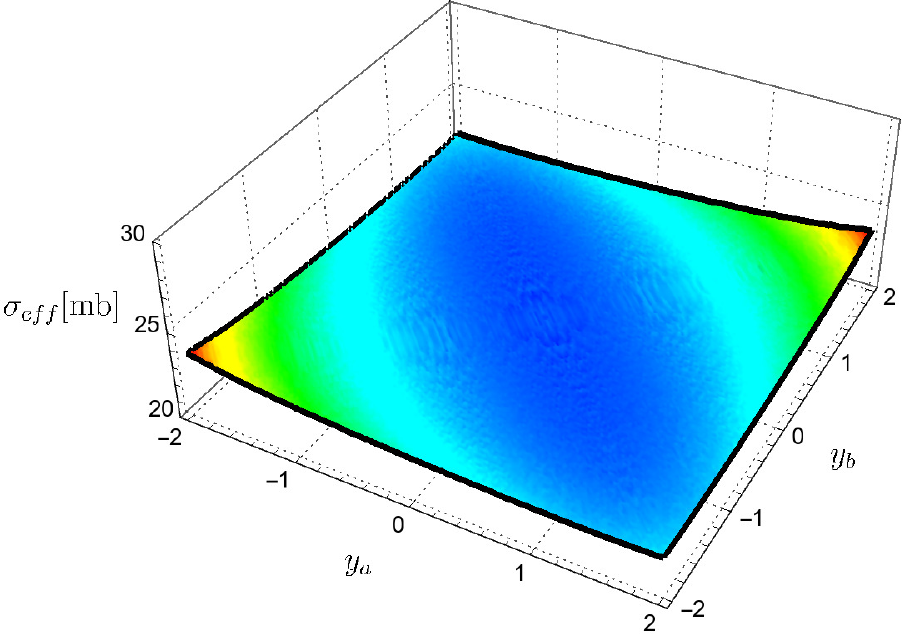}
\hskip 1cm
\includegraphics[scale=0.6]{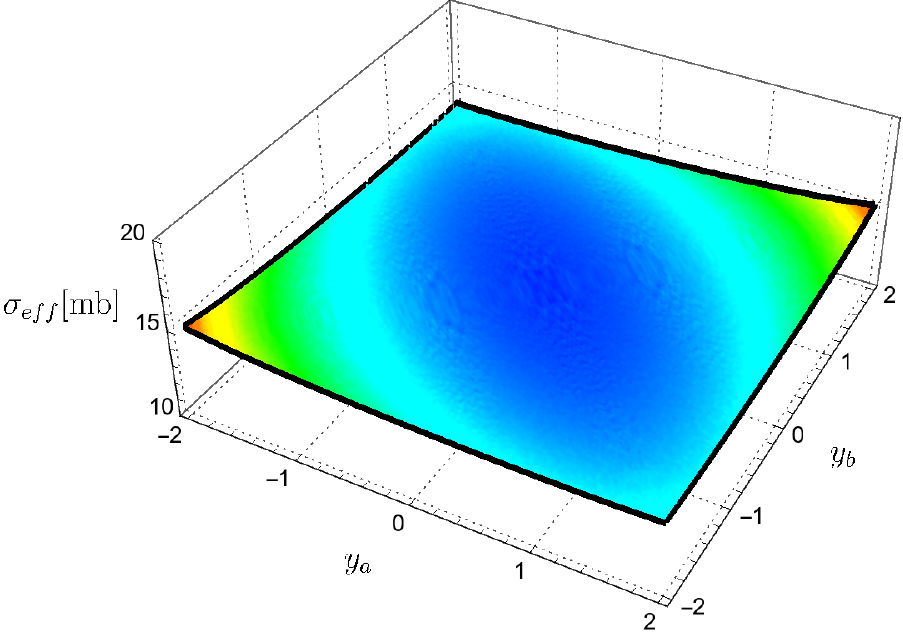}
\vskip 0.1cm
\caption{ \textsl{Effective cross section for the process $pp\rightarrow 
H H X$ as a function of Higgs bosons rapidities, $y_a$ and $y_b$, 
respectively in the central rapidity region. $\sigma_{eff}$
has been evaluated by using digluon distributions
at the scale $Q^2=m_H^2$. In the upper panels results
are shown  within the HO model with (left) and without (right) 
Melosh operators.
In the lower panels 
 results are shown for the HP model with (left) 
and without (right) Melosh operators.}}
\label{f_shCMS}
\end{figure*}
\begin{figure*}[h]
\centering
\includegraphics[scale=0.6]{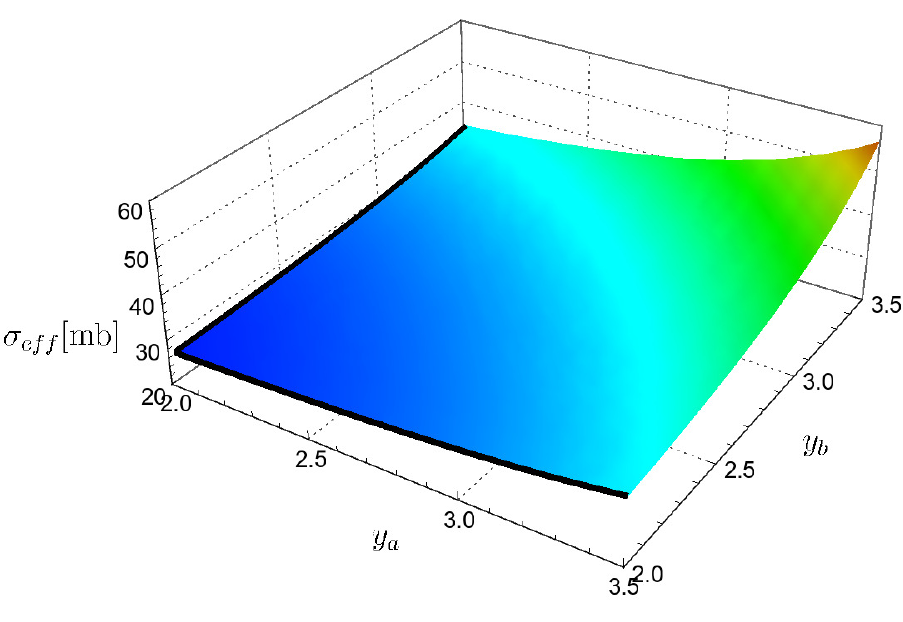}
\hskip 1cm
\includegraphics[scale=0.6]{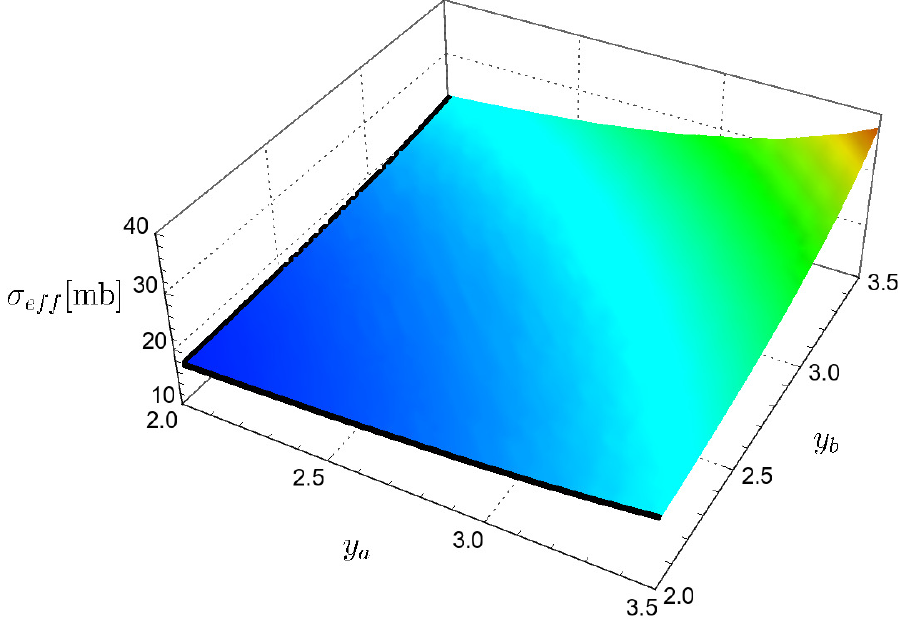}
\vskip 0.1cm
\includegraphics[scale=0.6]{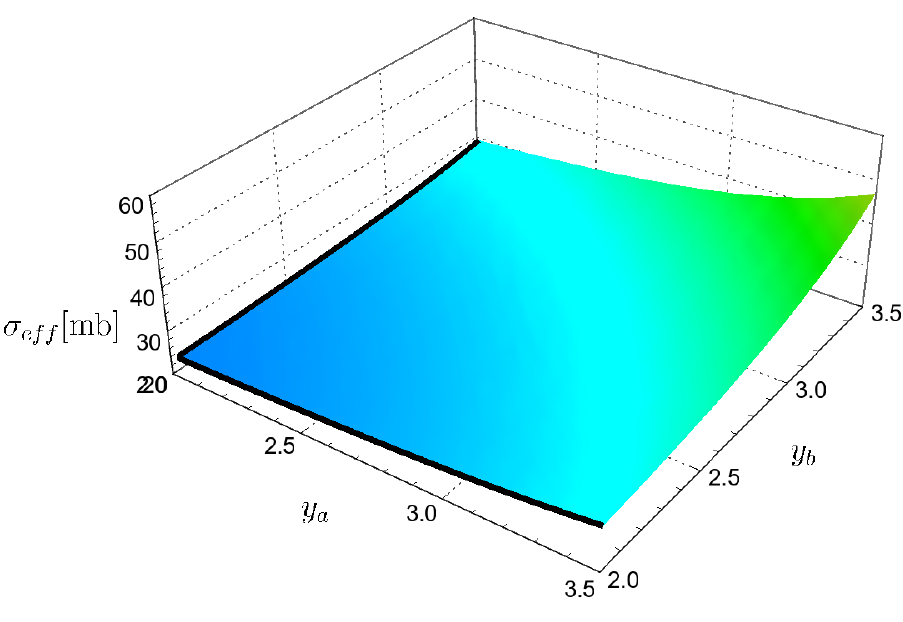}
\hskip 1cm
\includegraphics[scale=0.6]{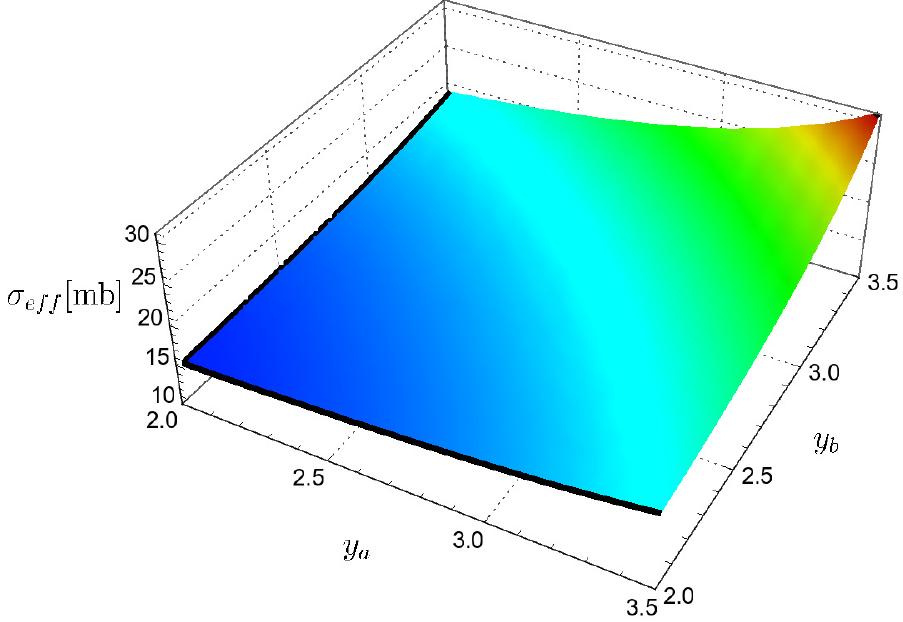}
\vskip 0.1cm
\caption{ \textsl{The same as in Fig. \ref{f_shCMS} but in the forward 
rapidity region.}}
\label{f_shLHCb}
\end{figure*}
\begin{figure*}
\centering
\hskip -1.3cm
\includegraphics[scale=0.450]{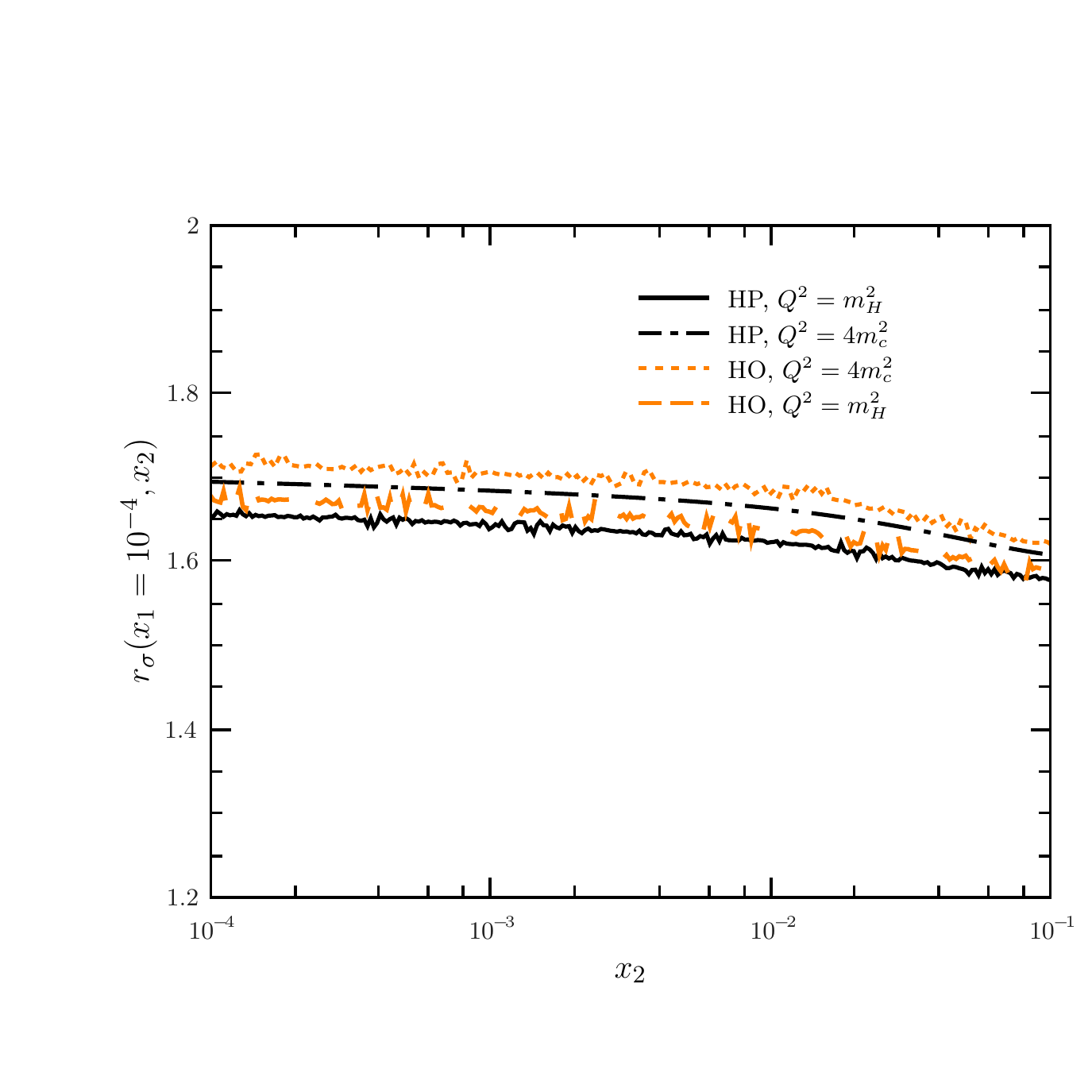}
\hskip-0.5cm
\includegraphics[scale=0.450]{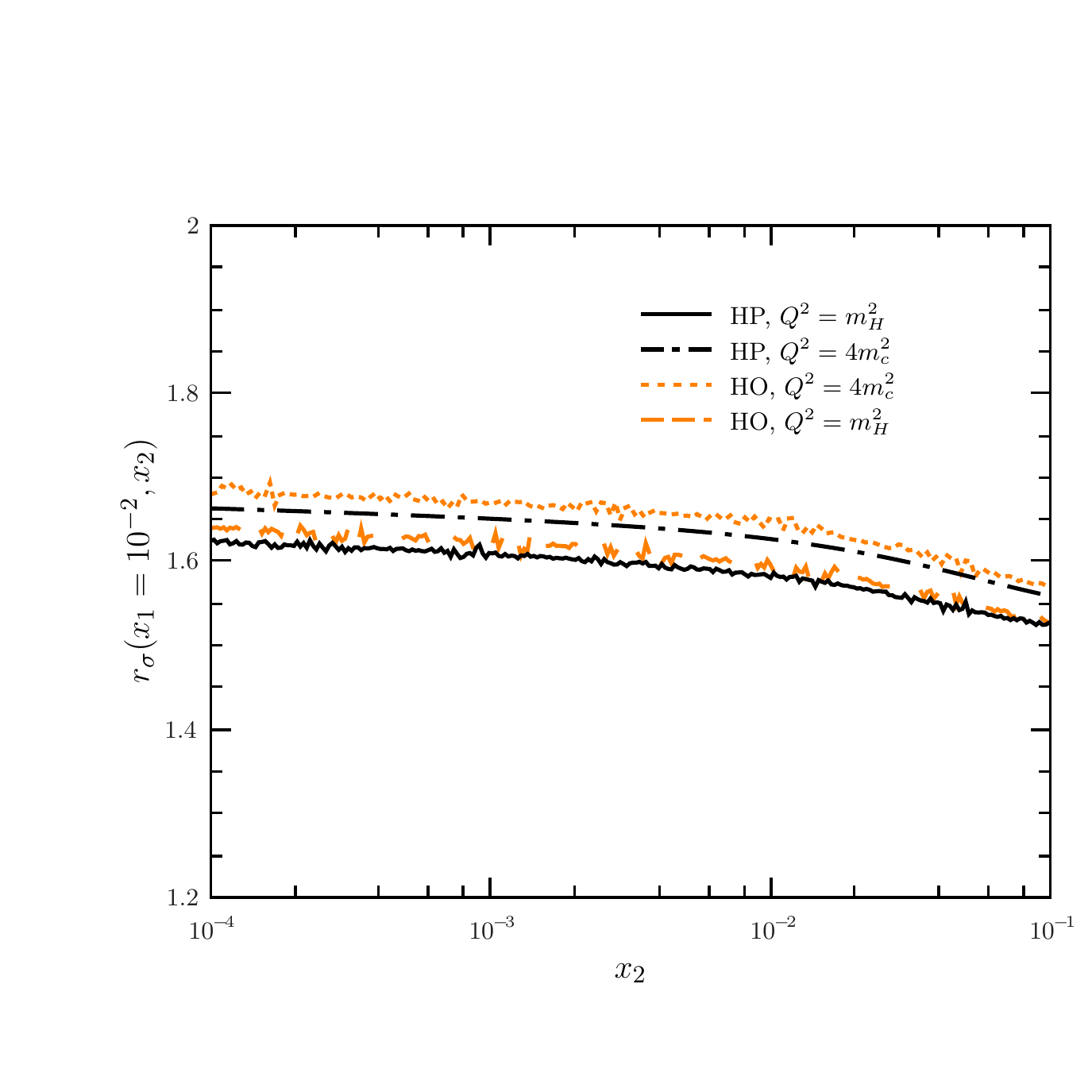}
\hskip-0.5cm
\includegraphics[scale=0.450]{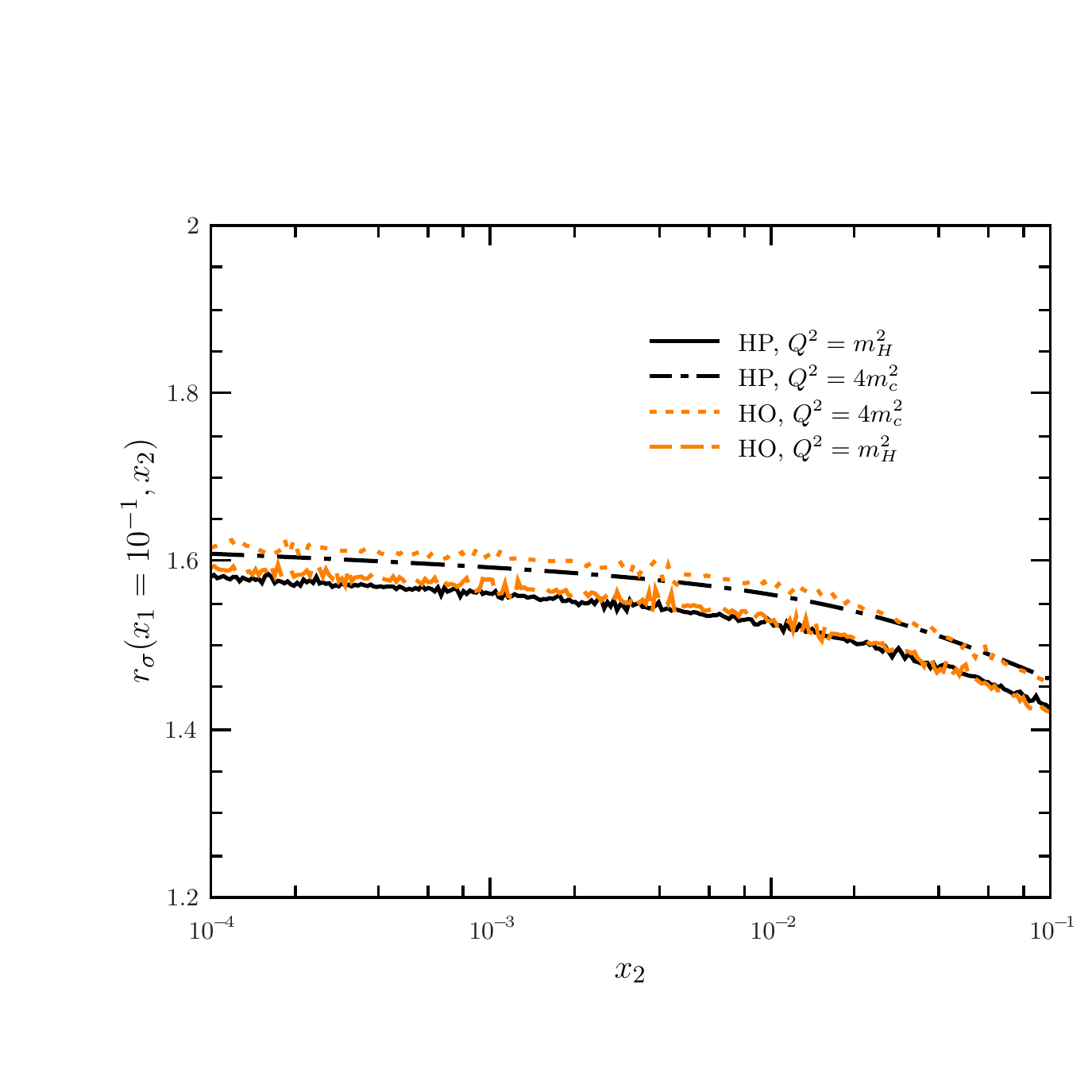}
\caption{ \textsl{The ratio in Eq. (\ref{rsi}) as a function of $x_2$ 
evaluated at fixed $x_1=10^{-4}$ (left panel),  $x_1=10^{-2}$ (central panel) 
and $x_1=10^{-1}$ (right panel).  This quantity
has been calculated via the digluon distribution computed within the HP
(full line) and the HO (dashed line) models at $Q^2=m_H^2$.
The same quantity is shown within the HP (dot-dashed line) and the HO 
(dotted line) models at $Q^2=4 m_c^2$.}}
\label{f_rs2}
\end{figure*}

\noindent
In Fig. \ref{tomo1} we present the double gluon distribution in 
coordinate space, $\tilde F_{gg}(x_1,x_2, b_\perp,Q^2=m_H^2)$, evaluated 
in different configurations of $x_1$ and $x_2$, including (black full 
lines) and neglecting (orange dashed lines) Melosh operators within 
different hadronic models, the HP in the upper panels and the HO in the 
lower ones. Results are consistent with those of Ref. \cite{noir}, where 
only valence quark dPDFs have been evaluated at the low hadronic scale of the 
models. 
We observe in the plots that there exists a value of $b_0\sim 1.5$ 
Ge$\mbox{V}^{-1}$, which slightly depends upon the kinematics and the 
hadronic model used in the calculations, such that the inclusion of 
Melosh operators strongly decrease dPDFs for $b_\perp<b_0$ and slightly 
increase them for $b_\perp>b_0$. 
It is worth noticing that Melosh operators 
reduce to the identity for $k_\perp=0$, so that dPDFs with and without 
Melosh coincide in this limit. Since the latter condition corresponds to 
an integral of dPDFs over $d^2 b_\perp$, it follows that dPDFs with and 
without Melosh are
normalized to the same number.
The digluon distributions in Fig. \ref{tomo1} show a marked dependence 
on the specific proton wave function built-in the CQMs. It is therefore  
instructive to present the ratio:
\begin{align}
r_b(x_1,x_2,b_\perp,Q^2) = { \tilde F_{gg}(x_1,x_,b_\perp,Q^2) \over 
\tilde F_{gg}^{NM}(x_1,x_2,b_\perp,Q^2) }~,
\label{r_rb}
\end{align}
where we indicate with $\tilde F^{NM}_{gg}(x_1,x_2, b_\perp, Q^2)$ the 
dPDFs in Eq. (\ref{dpdf}) evaluated neglecting Melosh operators. 
The ratio in Eq. (\ref{r_rb}) is shown in Fig. \ref{tomo2} with 
calculations 
performed within the HP model 
at the final scales $Q^2=m_H^2$ (full lines) and $Q^2=4m_c^2$ 
(dot-dashed lines), and 
the HO model at the final scales $Q^2=m_H^2$ (dashed lines) and 
$Q^2=4m_c^2$ 
(dotted lines) in three  configurations of $x_1$ and $x_2$. As one can 
see, up to $b_\perp < b_0$, Melosh operators 
induce a sizeable reduction of dPDFs which 
is almost a kinematical and scale independent effect. These conclusions 
hold for both the considered CQMs, which give rather close results. 
It is also interesting to study the impact 
of Melosh rotations directly 
on experimental related observables, such as $\sigma_{eff}$.
We first consider the production 
of double $J/\Psi$ via DPS at the LHC. Calculations are performed in the 
rapidity range $|y|<1.2$ for ATLAS and CMS kinematics and  $2<y<4.5$ for 
the LHCb one.
The calculation of $\sigma_{eff}$ is performed via digluon distribution 
evaluated at $Q^2=4m_c^2$.
In both these rapidity ranges, the involved parton momenta are quite 
small and 
we found that $\sigma_{eff}$ is nearly constant. For this reason we just 
quote the averaged results in Tab. \ref{tablesigma}.
The inclusion of Melosh operators determines an increase in 
$\sigma_{eff}$ by almost $60\%$, whereas there is only a slight 
dependence on the chosen hadronic model.
Then we consider double Higgs production via DPS in the same kinematic 
range. In this case, the digluon distribution is evaluated at 
$Q^2=m_H^2$. The results for $\sigma_{eff}$, as a function of final 
state particle rapidities, are shown in Figs. \ref{f_shCMS} and 
\ref{f_shLHCb}.
We note that $\sigma_{eff}$ is almost constant in the central rapidity 
region, as already observed at $Q^2=4 m_c^2$. However, for $Q^2=m_H^2$ 
in LHCb kinematics, the involved $x_i$ are substantially  higher with 
respect to those addressed in the $Q^2=4m_c^2$ case and $\sigma_{eff}$ 
starts to show a non trivial $x$ dependence.
From this plots it is clear that the production of heavy particles in the 
forward rapidity region represents a way to access the kinematic region 
where longitudinal correlation are the strongest.
For both the considered final scales, the inclusion of Melosh operators 
increase the value of $\sigma_{eff}$, as they act to reduce the size of 
dPDFs at small $b_\perp$, as shown in Fig \ref{tomo1}. 
\begin{table}[t]
\begin{center}
\centering
\begin{tabular}{c|c|c|c|c|c|c|c|c|c|c|c|} 
\cline{1-9}             
\multicolumn{3}{|c|}{Kinematics}&\multicolumn{3}{c|}{HO 
model}&\multicolumn{3}{c|}{HP 
model} \\ 
\cline{4-9}
\multicolumn{3}{|c|}{$Q^2 = 4 m_c^2$}&\multicolumn{2}{c}{ 
$\sigma_{eff}$ [mb] }&\multicolumn{1}{c|}{ 
$\sigma_{eff}^{NM}$ [mb]}
&\multicolumn{2}{c}{$\sigma_{eff}$ [mb]}&
\multicolumn{1}{c|}{$\sigma_{eff}^{NM}$ [mb]} \\ 
\cline{1-9}
\multicolumn{3}{|c|}{$|y| < 1.2$  
} &  \multicolumn{2}{c}{23.5}    & 
\multicolumn{1}{c|}{13.7}&  \multicolumn{2}{c}{21.0}    & 
\multicolumn{1}{c|}{12.4} \\
\multicolumn{3}{|c|}{ $2<y<4.5$ } 
&  \multicolumn{2}{c}{23.6}   & 
\multicolumn{1}{c|}{13.9}&  \multicolumn{2}{c}{21.1}   &  
\multicolumn{1}{c|}{12.6}  \\ 
\hline
\end{tabular}
\caption{ \textsl{Calculations of $\sigma_{eff}$  in the relevant 
experimental rapidity range of the process $pp\rightarrow J/\Psi J/\Psi 
X$. Results are presented for digluon 
distribution evaluated at $Q^2 = 4 m_c^2$ and
obtained within the HO and HP models, including and neglecting Melosh 
operators.}}
\label{tablesigma}
\end{center}
\end{table}  
The above results are similar, in quality, to that discussed in 
Ref. \cite{noiplb1}. 
In order to further explore the role of Melosh operators in 
$\sigma_{eff}$, we consider the following ratio \cite{noir}:
\begin{align}
\label{rsi}
 r_\sigma(x_1,x_2) = \dfrac{ 
\sigma_{eff}(x_1,x_2)}{\sigma^{NM}_{eff}(x_1,x_2)}~,
\end{align}
where in the denominator the effective cross section has been evaluated by 
means of gluon dPDFs calculated without Melosh rotations. Results of numerical 
calculations are presented in Fig. \ref{f_rs2} for three fixed typical 
values of $x_1$.  
Such a ratio shows a very weak dependence on $x$ and the chosen model,
and a weak dependence on the hard scale. Moreover its numerical value is  found 
to be quite close to that obtained  with valence quarks dPDFs evaluated
at the hadronic scale of models described in Ref.~\cite{noir}. It is 
interesting to note that Melosh's effects on 
$\sigma_{eff}$ by far exceed the dependence induced by using different 
hadronic models.

\label{paras}
As already discussed, Melosh operators  encode $x-k_\perp$ correlations which guarantee the 
frame independence of the Light-Front wave function, 
an essential property which dPDFs must satisfy too. 
As previously shown above, Melosh effects  on dPDF 
calculations are rather independent with respect to the adopted CQM, see Fig. 
\ref{tomo2}  and Ref. \cite{noir}. 
Moreover, by 
comparing Figs. \ref{tomo2} and \ref{f_rs2} with the corresponding 
Figs. 7 and 8 of Ref.~\cite{noir}, one may notice that  such effects are also rather independent on the flavor of the active partons.
In addition, as shown in Figs. \ref{tomo2},\ref{f_rs2}, Melosh  
effects  {mildly}  depend on typical  scales involved 
in the
hard scatterings, either the $J/\Psi$ or the Higgs mass in the present analysis.
{These features suggest that one may study the functional form of these $x-k_\perp$ correlations which can then be used to inspire dPDFs phenomenological models.}
For this purpose 
we define the ratio $R$ between digluon PDFs calculated within CQM 
in a fully LF calculation and its approximation obtained neglecting 
 Melosh operators:
\begin{align}
R(x_1,x_2,k_\perp) =  {
  F_{gg}(x_1,x_2,k_\perp, Q^2) \over 
 F_{gg}^{NM}(x_1,x_2,k_\perp,Q^2) }
~.\label{rhp}
\end{align}
Such a ratio is built in order to suppress dynamical effects 
encoded in the chosen hadronic wave function.
In fact, since we are interested  in $x-k_\perp$ correlations
induced only by Melosh operators, we have evaluated the ratio in Eq. (\ref{rhp})
within the only model which does not include any additional 
$x-k_\perp$ correlation generated by its wave function, \textsl{i.e.} the HO 
model \cite{noiold,noir}. 
\begin{figure*}
\centering
\includegraphics[scale=0.55]{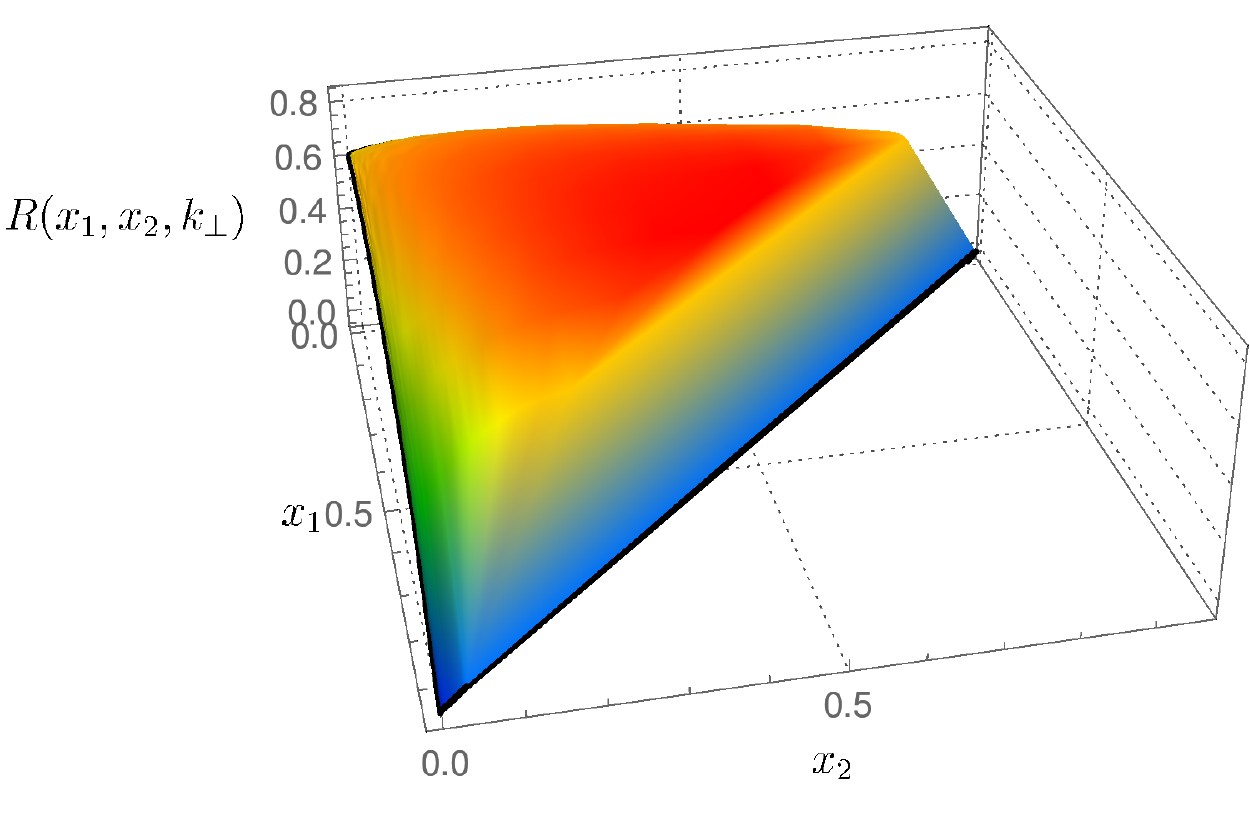}
\includegraphics[scale=0.55]{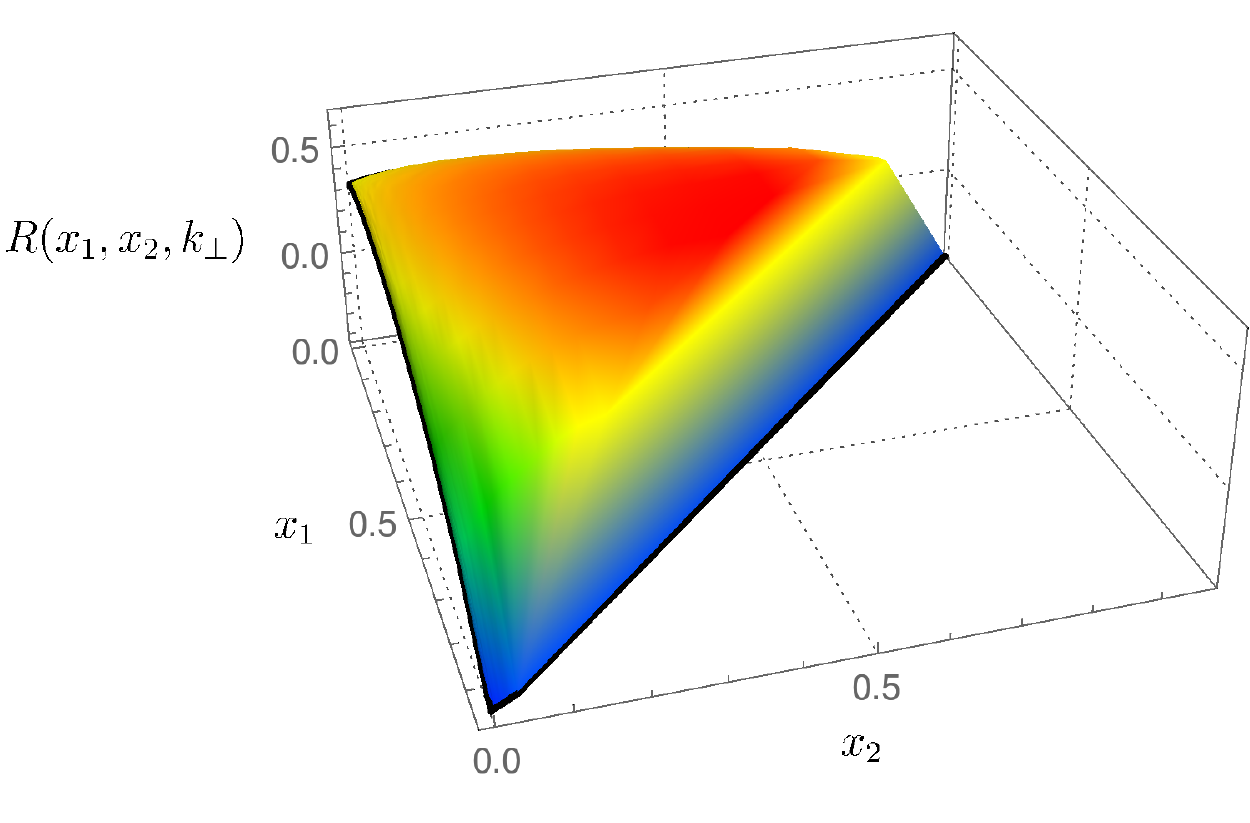}
\caption{ \textsl{The ratio $R_{gg}(x_1,x_2,k_\perp)$ evaluated 
within the HO model at the final scale $Q^2=m_H^2$. Left panel for 
$k_\perp=0.15$ GeV. Right panel for $k_\perp=1.97$ GeV. } }
\label{fit4b}
\end{figure*}
We display in Fig. \ref{fit4b} the ratio $R(x_1,x_2,k_\perp)$
for two representative values of $k_\perp$
as a function of $x_1$ and $x_2$. 
We found that a  suitable 
parametrization in $x_1-x_2$ space, 
able to describe the ratio $R$ at fixed $k_\perp$, is the following one:
\begin{align}
R(x_1,x_2, k_\perp) &= 
w( k_\perp)
(x_1 x_2)^{t( k_\perp)}
(1-x_1-x_2)^{|x_1-x_2|e( k_\perp) 
} e^{-(1-x_1-x_2)h( k_\perp)}~.
\label{r_fit1}
\end{align}
The parameter $w$ controls the overall normalization of $R$, $t$ its small-$x$ behaviour. {The additional parameter $e$ and $h$ control its behaviour
on the $x_1+x_2=1$ boundary}. 
Such a functional form goes beyond the standard factorized
ansatz often used for dPDFs.
By using the functional form in Eq. (\ref{r_fit1}), we perform a series of 
fit of $R(x_1,x_2,k_\perp)$ at fixed values of $k_\perp$. This procedure gives 
us access to the $k_\perp$ dependence of the parameters which 
is displayed in Fig. \ref{fit2a}.
Then, the $k_\perp$ dependence is interpolated by a fourth order polynomial of the type:
\begin{align}
i(k_\perp) =d_i+ a_i k_\perp^2 + b_i  k_\perp^3 + c_i k_\perp^4~, \;\; i=\{w,e,t,h\},
\label{r_fit2}
\end{align}
involving four parameters for each $i$. 
For $k_\perp=0$ Melosh operators reduce to unity, $R(x_1,x_2,k_\perp=0)=1$, and therefore 
$e(k_\perp=0)=h(k_\perp=0)=t(k_\perp=0)=0$. The latter condition is  fulfilled 
by setting $d_e=d_h=d_t=0$ and $d_w=1$ which are held fixed at those values 
during the fit. 
The corresponding results are displayed as solid lines in Fig. \ref{fit2a} 
and the best fit parameters are reported in Table \ref{t3}. 
It is worth noticing that $w$ is compatible with unity and that $t$ 
is compatible with zero: Melosh operators mainly affect the behaviour of
the ratio on the kinematic boundary. 
The dampening on the boundary is increasingly pronounced as $k_\perp$ increases. 
The obtained parametrization
reproduces with good accuracy
(at the percent level)
the ratio $R(x_1,x_2,k_\perp)$ calculated within the HO model. 
Additionally, its investigation at different scales, reveals that 
is substantially scale independent.

\begin{figure}
\centering
\includegraphics[scale=0.6]{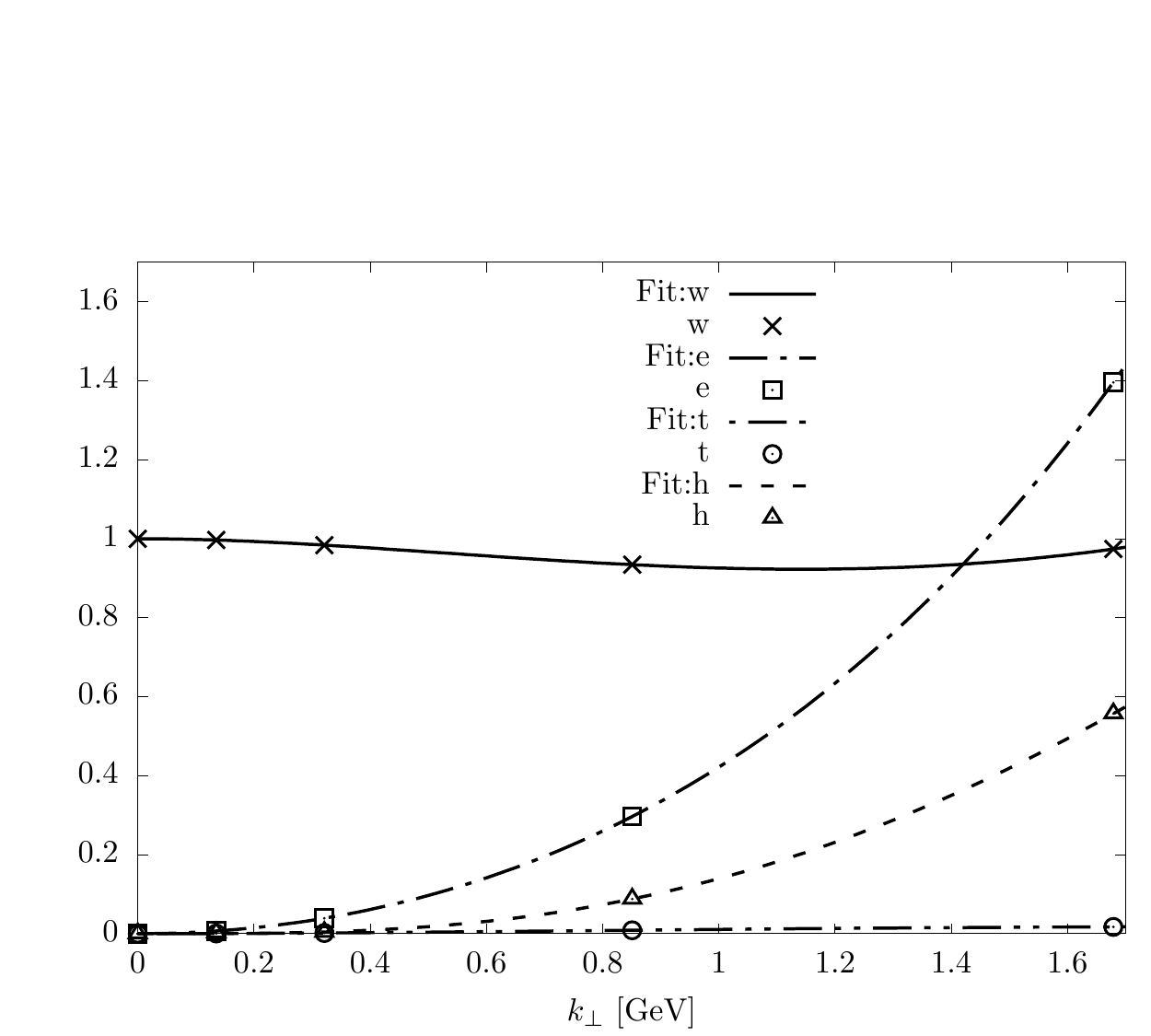}
\caption{ \textsl{Symbols indicate the values of the parameters
$w,e,t,h$ obtained by fitting the ratio R with Eq. (\ref{r_fit1}) 
for fixed values of $k_\perp$.
Lines indicate fit to these points obtained by using the functional 
form Eq. (\ref{r_fit2}).}}
\label{fit2a}
\end{figure}

\begin{table}[t]
\centering
\begin{tabular}{|c|c|c|c|c|c|}
\hline
$i$  & $a_i$ & $b_i$ & $c_i$ & $d_i$\\
\hline
$w$ & -0.202 & 0.146 & -0.019 & 1\\ 
$e$ & 0.369 & 0.019 & 0.033 & 0\\
$t$ & 0.021  &-0.012  &0.002  & 0\\
 $h$ & -0.019 & 0.202 &-0.043 & 0\\ 
 \hline
\end{tabular}
\caption{ \textsl{Values of the coefficients in the parametrizations in Eq.
(\ref{r_fit2}) as returned by the fit to the $k_\perp$ dependence of the ratio 
Eq. (\ref{rhp}) evaluated within the HO model. The $d_i$ values are held fixed 
during the fit.}}
\label{t3}
\end{table}

\section[Conclusions]{Conclusions}
\label{conc}
\noindent
In the present analysis we have  investigated to which extent information on
the partonic proton structure, complementary to that 
obtained via other parton distributions, can be accessed via dPDFs. In 
particular we have focused our attention on the connection between the 
mean transverse partonic distance between two partons  and  $\sigma_{eff}$. 
We have 
discussed  how this relation is modified when correlation of 
perturbative and non-perturbative origin are included in the 
calculation. In the former we have considered perturbative correlations 
induced by the so called splitting term in dPDF evolution. 
In the latter we have considered 
non perturbative correlation 
beyond the factorized ansatz for dPDFs.
We proved that also in these two cases, the mean value of $\sigma_{eff}$ 
provides new indications on the structure of the proton in the non perturbative 
regime of QCD, again indicating dPDFs as a valuable tool to investigate partonic 
longitudinal and transverse  correlations. In the last part of this work  we 
took advantage of CQM calculations of dPDFs 
within the Light-Front relativistic approach, to study model independent 
correlations between $x_1,x_2$ and $k_\perp$ induced by the so called Melosh 
operators.  We have investigated
their effects on the digluon dPDF, perturbatively obtained at high momentum 
scales relevant  for DPS studies at the LHC. 
We have shown that Melosh operators produce a non-negligible reduction of 
dPDFs and generate, model independent, $x_i-k_\perp$ correlations
on the kinematic boundary.
 

 

\end{document}